\documentclass[aps,pre,preprint]{revtex4}

\usepackage{amsmath,epsfig,amsfonts,epsfig,graphicx,psfrag,natbib
}

\begin{document}

\title{Community Structure in the United States House of Representatives}

\author{Mason A. Porter}
\affiliation{Department of Physics and Center for the Physics of Information, California
Institute of Technology, Pasadena, CA 91125, USA}
\author{Peter J. Mucha}
\affiliation{Department of Mathematics and Institute for Advanced Materials, University of North Carolina, Chapel Hill, NC  27599-3250 }
\author{M. E. J. Newman}
\affiliation{Department of Physics and Center for the Study of Complex Systems, University of Michigan, Ann Arbor, MI  48109-1120}
\author{A. J. Friend}
\affiliation{School of Mathematics, Georgia Institute of Technology, Atlanta, GA  30332-0160}



\begin{abstract}
  We investigate the networks of committee and subcommittee assignments in
  the United States House of Representatives from the 101st--108th
  Congresses, with the committees connected by ``interlocks'' or
  common membership.  We examine the community structure in these networks
  using several methods, revealing strong links between certain committees
  as well as an intrinsic hierarchical structure in the House as a whole.
  We identify structural changes, including additional hierarchical levels
  and higher modularity, resulting from the 1994 election, in which the
  Republican party earned majority status in the House for the first time
  in more than forty years.  We also combine our network approach with
  analysis of roll call votes using singular value decomposition to uncover
  correlations between the political and organizational structure of House
  committees.
\end{abstract}

\maketitle





\section{Introduction}

Much of the detailed work in making United States law is performed by
Congressional committees and subcommittees.  This contrasts with
parliamentary democracies such as Great Britain and Canada, in which a larger
part of the legislative process is directly in the hands of political
parties or is conducted in sessions of the entire parliament.  While the
legislation drafted by committees in the U.S.~Congress is subject
ultimately to roll call votes by the full Senate and House of
Representatives, the important role played by committees and subcommittees
makes the study of their formation and composition vital to understanding
the work of the American legislature.

The presence of committees in the House endows it with obvious hierarchical
levels: individual Representatives, subcommittees, standing committees, and
the entire House floor.  However, it is desirable to examine social
networks in the House of Representatives quantitatively to determine
whether it has any additional structure that might relate to collaborative
efforts among Representatives, such as correlations or close associations
between the members of different committees.  The importance of such
studies is not merely academic.  An understanding of the House as a
collaboration network may shed considerable light on the law-making
process, as bills often spend time in several different committees and
subcommittees while being drafted in preparation for votes on the House
floor.  For instance, the House's consideration of the No Child Left Behind
Act of 2001 involved the Committee on Education and the Workforce, the
Subcommittee on Education Reform, the Subcommittee on 21st Century
Competitiveness, the Committee on the Judiciary, and the Committee on Rules
(setting the terms for the scrutiny of the bill) before being approved by
the full House~\cite{thomas}.  After the Senate further amended the bill, a
conference agreement eventually passed both houses of Congress and the
final bill was signed by the President to become public law No.~107--110.

Analyzing the structure of the committee system in the House of
Representatives and studying its correlation with the partisanship of its
constituent Representatives helps achieve a better understanding of
political party competition in Congress.  Several contrasting theories of
committee assignment have been developed in the political science
literature (mostly through qualitative studies, although there have been
some quantitative ones; see \cite{nisk,gill,kreh,cox,shep,org}), but there
is no consensus explanation of how committee assignments are initially
determined or how they are modified from one two-year term of Congress to
the next.  A question of particular interest is whether political parties
assign committee memberships essentially at random or if, for instance, one
can show using objective analyses that influential Congressional committees
are ``stacked'' with partisan party members.

\begin{figure}
  \centerline{(a) \includegraphics[width=0.4\textwidth]{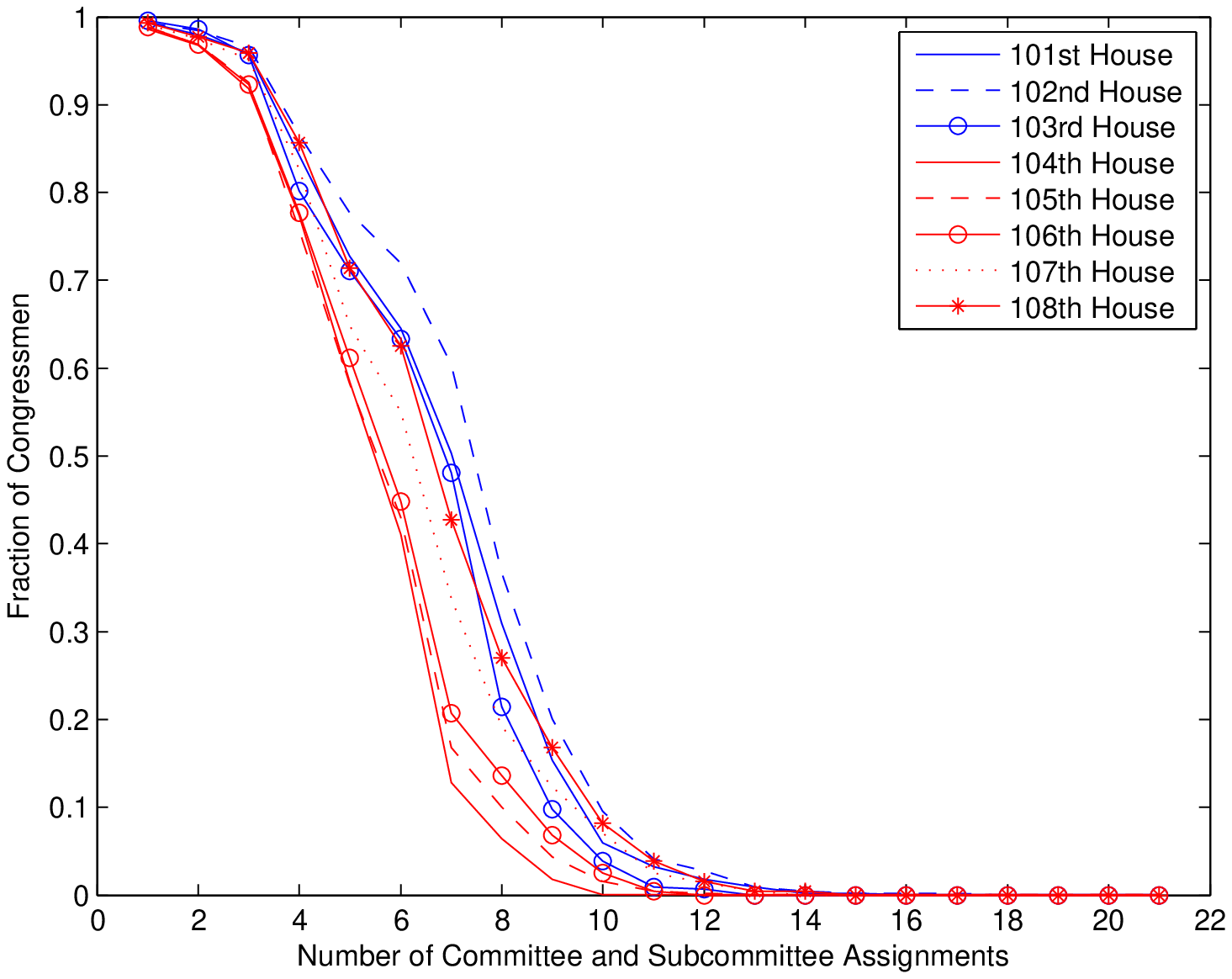} \hspace{1cm}
  (b) \includegraphics[width=0.4\textwidth]{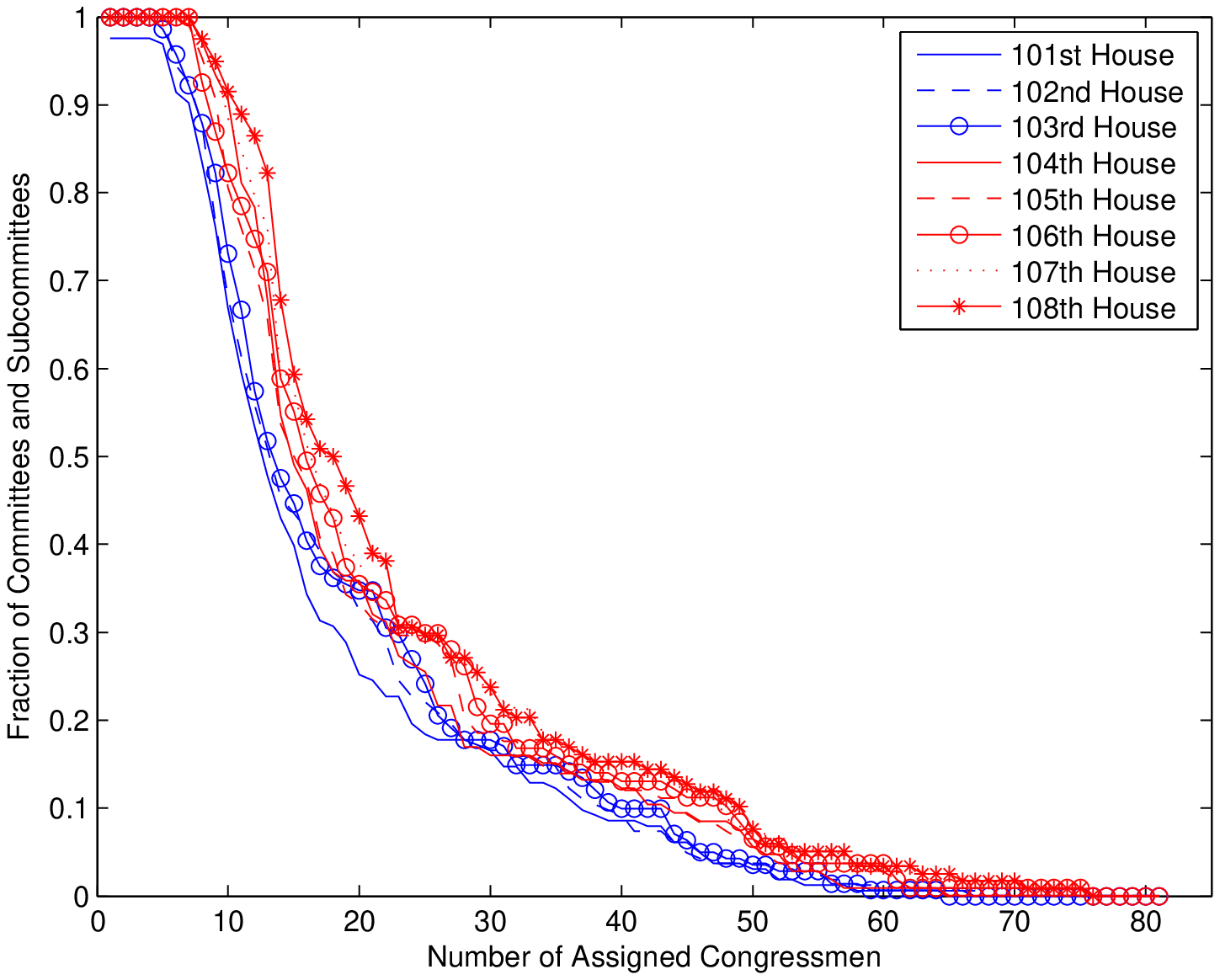}}
\caption{(Color) Cumulative degree distributions of the 101st--108th
  U.S.~House of Representatives networks defined by committee and
  subcommittee assignments.  (a) Fraction of Representatives versus number
  of (sub)committee assignments.  (b) Fraction of committees and
  subcommittees versus number of assigned Representatives.  In the
  104th--108th Houses, all with Republican majorities, the cumulative
  degree distribution in (a) shifts farther up in each House and that in
  (b) shows a similar but less pronounced shift.  There is no noticeable
  trend in the Democrat-majority 101st--103rd Houses in (a), but it seems
  to shift up a bit in (b) to reveal a drift covering all eight
  Congressional terms we studied.}
\label{revo}
\end{figure}

Our study of the organizational structure of Congress draws on network
theory, which provides powerful tools for representing and analyzing
complex systems of connected agents.  While the quantitative study of
real-world networks has a long history in the social sciences (see, for
example, the discussions in \cite{newmansirev,wattsreview}), such
investigations experienced a major expansion in popularity in the late
1990s, in part because of interest in the Internet and online networks.
This increased attention has been especially evident in studies of large
social, biological, and technological networks, which have relied on major
advancements in computer hardware and algorithms to generate novel results
\cite{str01,albert,newmansirev,wattsreview,mendes}.  Among the myriad
topics that have been studied are evolving social groups \cite{kossin},
collaborations \cite{ama}, community detection \cite{commreview}, and
hierarchical organization \cite{ravasz,ama07}.  It is the modular and
hierarchical community structure of networks that primarily concern us in
our present study of Congressional committee assignments.

The Congressional networks studied here are examples of collaboration networks, on which there is a considerable body of previous literature.  Networks constructed from collaborations between
corporate boards of directors~\cite{Mariolis75,Useem84,schwartz,mizruchi,robins} are especially germane to the present work, as such collaborations occupy
a position in the business world somewhat analogous to that of
collaborations between Congressional committee members.  Previous studies
have shown that board memberships play a major role in the spread of
attitudes, ideas, and practices through the corporate world, affecting
investment strategies \cite{Haunschild93}, political
donations~\cite{Useem84}, and even the stock market on which a company is
listed~\cite{RDW00}.  Studies of the structure of corporate networks have
shed light on the mechanisms and pathways of information
diffusion~\cite{ceo,burt2,burt}, and we believe that the structure of
congressional committees may turn out to be similarly revealing.

As we show here, network methods are particularly effective at uncovering
structure among committee and subcommittee assignments in Congress, without
the need to incorporate any specific knowledge about committee members or
their political positions.  In a recent article~\cite{congshort}, the
present authors formulated and briefly examined a number of committee
assignment networks, looking, for instance, at the partisanship of the
Select Committee on Homeland Security in the 107th House and its
connections to other committees.  An alternative network perspective on the
structure of Congress has been offered by J.~H.
Fowler~\cite{fowlershort,fowler}, who examined the network defined by the
cosponsorship of legislation by Members of Congress.  In the present work,
we compare our previous observations to those for the 108th House and
explore the structural changes in the networks that resulted from the 1994
Congressional elections in which the Republican party gained majority
control of the House.  A detailed technical discussion of the methods used
to obtain our results is included in the appendices.

This paper is organized as follows.  First, we define the bipartite
collaboration networks determined by the assignments of Representatives to
House committees and subcommittees.  We then investigate the hierarchical
and modular structure of these networks using several different community
detection methods.  We also incorporate singular value decomposition (SVD)
analysis of House roll call votes into our study of the House's community
structure.  We provide additional details in two appendices: in Appendix~A,
we explain the methods used in our SVD analysis of voting patterns; in
Appendix~B, we give a detailed comparison of several methods for community
detection in networks, including our generalization of a recently proposed
local detection algorithm~\cite{bagrow} to weighted networks.

\section{Committee assignment networks in the House}
\label{networksec}

\begin{figure}
\begin{center}
\includegraphics[width = 14cm]{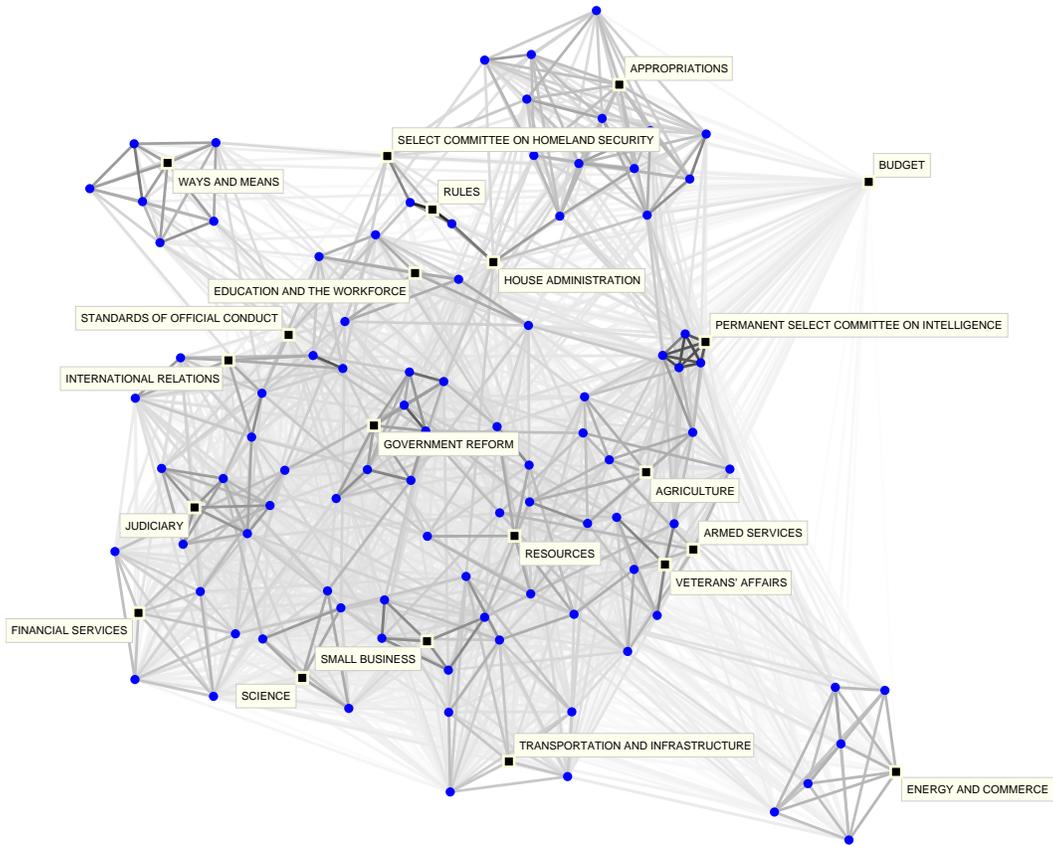}
\end{center}
\caption{(Color online) Network of committees (squares) and subcommittees
  (circles) in the 107th U.S.~House of Representatives, with standing and
  select committees labeled.  (Subcommittees tend to be closely tied to
  their main committee and are therefore left unlabeled.)  Each link
  between two (sub)committees is assigned a strength (indicated by the
  link's darkness) equal to the normalized interlock.  (The ``interlock"
  between two committees is equal to the number of their common members.
  The normalization takes committee sizes into account by dividing the raw
  interlock by the expected number of common members if assignments were
  determined independently and uniformly at random.)  Thus, lines between
  pairs of circles or pairs of squares represent normalized degree of joint
  membership between (sub)committees (it is because of this normalization
  that lines between squares are typically very light), and lines between
  squares and circles represent the fraction of standing committee members
  on subcommittees.  This figure is drawn using a variant of the
  Kamada-Kawai spring-embedding visualization, which takes link strengths
  into account~\cite{kk}.}
\label{network}
\end{figure}

\begin{figure}
\begin{center}
\includegraphics[width = \textwidth]{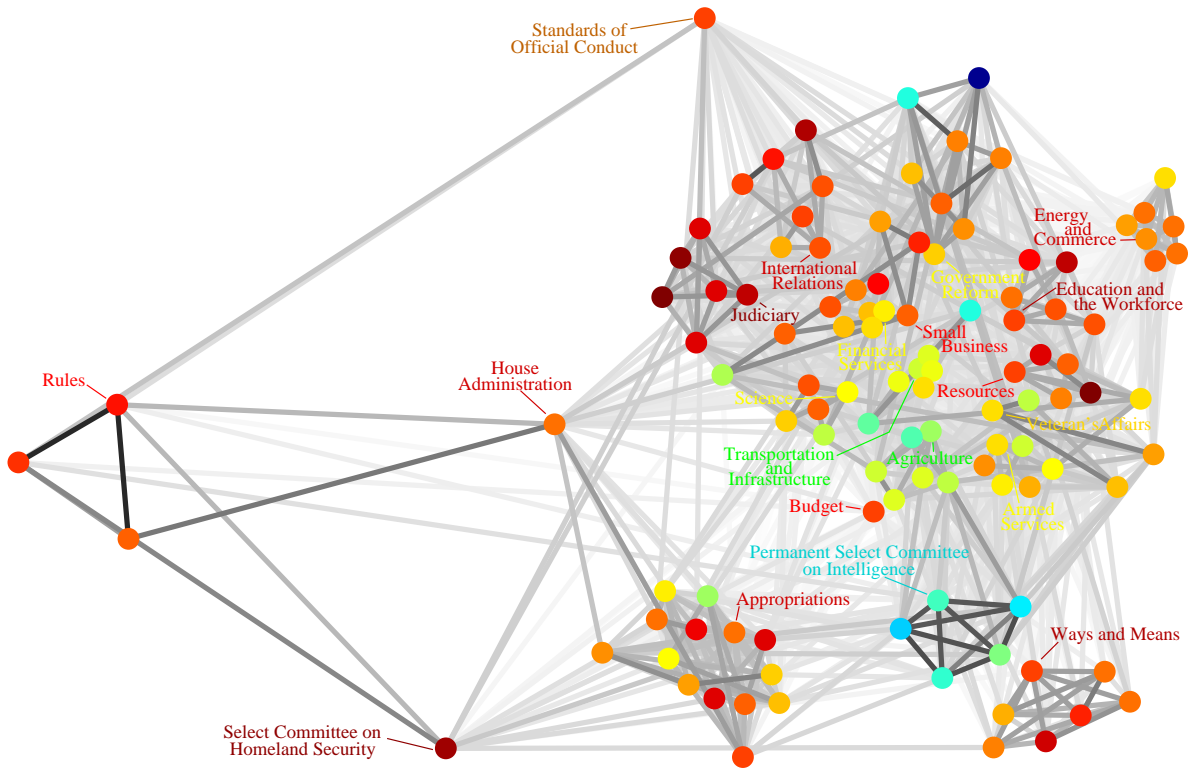}
\end{center}
\caption{(Color) Another visualization of the committee assignment network,
  with standing committees labeled.  This plot is color-coded according to
  the mean ``extremism" (defined in the text; see Appendix~A for details)
  of each committee's constituents, where red nodes are more extreme and
  blue ones are less extreme.  Thus, the redder committee are composed of
  ideologues from both parties and the bluer ones consist of moderates.
  The strength (darkness) of the link between two committees is equal to
  their interlock (which is not normalized here).  Observe that similar
  groups of committees are clustered together in this plot as in
  Fig.~\ref{network}, despite the different definition of link strengths.
  (This figure is also drawn using a variant of the Kamada-Kawai spring
  embedder.)}
\label{network2}
\end{figure}

\begin{figure}
\begin{center}
\includegraphics[width = \textwidth]{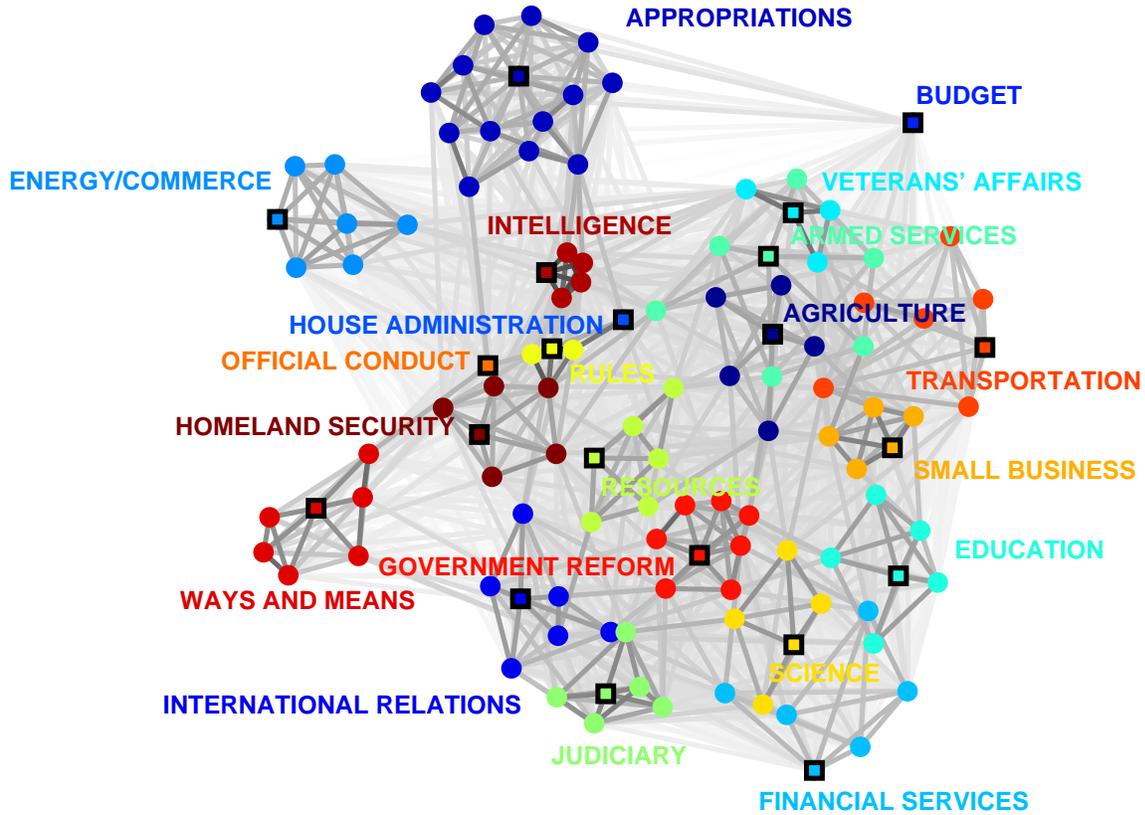}
\end{center}
\caption{(Color) Network of committees (squares) and subcommittees
  (circles) in the 108th U.S.~House of Representatives, color-coded by the
  parent standing and select committees.  (The depicted labels indicate the
  parent committee of each group but do not identify the location of that
  committee in the plot.)  As with Fig.~\ref{network}, this visualization
  was produced using a variant of the Kamada-Kawai spring embedder, with
  link strengths (again indicated by darkness) determined by normalized
  interlocks.  Observe again that subcommittees of the same parent
  committee are closely connected to each other.}
\label{network108}
\end{figure}

\begin{figure}
\begin{center}
{\includegraphics[width = 14cm]{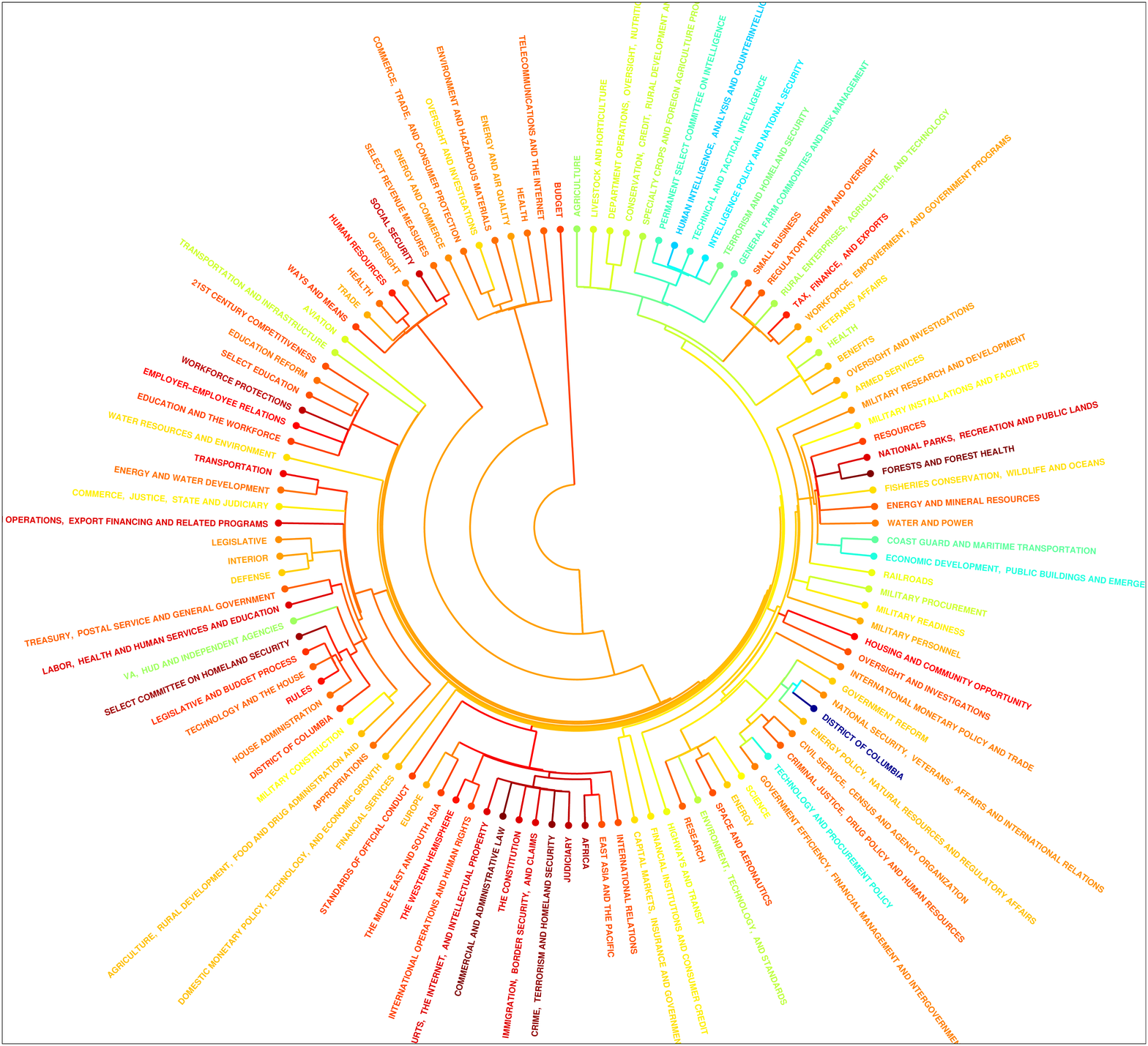}}
\end{center}
\caption{(Color) Dendrogram representing the hierarchical clustering of the
  committees in the 107th U.S.~House of Representatives, determined by
  single linkage clustering of normalized committee interlocks.  Each
  committee is color-coded according to the mean ``extremism'' of its
  members (defined in the main text; see Appendix~A), from less extreme
  (blue) to more extreme (red).  The clusters at each level are color-coded
  according to the average of their constituent committee extremism
  scores.}
\label{107clust}
\end{figure}

We represent each of the 101st--108th terms of the U.S.~House of
Representatives as a separate bipartite (two-mode) network based on
assignments of Representatives to committees and subcommittees (henceforth
called just ``committees'' for simplicity).  The two types of nodes in
these networks correspond to Representatives and committees, with edges
connecting each Representative to the committees on which he or she sits.
The period we study (1989--2004) spans the political changes following the
terrorist attacks of September 11, 2001~\cite{thomas}, as well as the 1994
elections in which the Republican party won majority control of the House
for the first time in more than forty years.  We construct one network for
each two-year Congressional term from data published by the Clerk's Office
of the House of Representatives~\cite{clerk}, ignoring changes in committee
assignments that occur while a term is underway.

Each network includes roughly 440 Representatives (including non-voting
Delegates and midterm replacements), about $20$ standing committees, and
more than $100$ subcommittees, with an average of about $6$ committee
assignments per Representative.  Because of the relatively high edge
density (about $5\%$ of possible connections are present), some frequently
studied network statistics, such as geodesic path lengths, turn out to be
unrevealing in this case.  Therefore, we instead focus our attention on
the community structure of the networks and associated measures such as
modularity and Horton-Strahler numbers.  We discuss these analyses in
Section~\ref{structure}.

With data for eight consecutive Congresses, it is natural to ask how the
committee assignment network changes in time.  One question of interest is
whether the networks contain signs of the so-called ``Republican
Revolution'' of 1994 that ended forty years of Democratic majorities in the
House of Representatives, the longest span of single-party rule in
Congressional history~\cite{revolution}.  That is, can one observe
structural differences in the committee assignment networks between the
Democrat-majority Houses (101st--103rd) and the Republican-majority ones
(104th--108th)?  As one means of addressing this question, we compute the
degree of each node, defined as the number of edges connected to it.
Because the committee assignment networks are bipartite, we construct two
types of cumulative (integrated) degree distributions~\cite{gf} and examine
how they changed across Congresses.  One distribution (Fig.~\ref{revo}a)
indicates the number of committees on which each Representative sits, and
the other (Fig.~\ref{revo}b) gives the number of Representatives on each
committee.  We do not observe a significant trend in Democrat-majority
Houses, although a slow increase in committee sizes is revealed in
Fig.~\ref{revo}b.  The committee reorganization that accompanied the
formation of the Republican-majority 104th House, however, produced a sharp
decline in the typical numbers of committee and subcommittee assignments
per Representative, but the trend in subsequent Republican-majority
Congresses has been a slow increase in both the numbers of assignments and
the committee sizes.  These trends are visible in Fig.~\ref{revo}.

While rich in their data content, the two-mode networks of committee
assignments are difficult to visualize and interpret.  A common strategy in
such cases is to examine instead a one-mode ``projection'' of the network
onto either the committees or the Representatives.  In our studies, we have
made considerable use of the projection onto the committees, in which a
network is created whose nodes represent the committees and whose edges
represent common membership or ``interlocks'' between committees.
Figures~\ref{network} and~\ref{network2} show two different visualizations
of the network of committees for the 107th House of Representatives
(2001--2002), an example that we analyze in some depth below.

We quantify the strength of a connection between committees in this
projected network with a \emph{normalized interlock}, defined as the number
of common members two committees have divided by the expected number of
common members if membership of committees of the same size were randomly
and independently chosen from Congressmen in the House.  Committees with as
many common members as expected by chance have normalized interlock~1,
those with twice as many have interlock~2, those with none have
interlock~0, and so forth.  We use this weighting in the visualization of
the network of the 107th House in Fig.~\ref{network} by darkening the links
between nodes accordingly \footnote{The Euclidean distances in these
  figures arise from a Kamada-Kawai force-directed network visualization
  algorithm \cite{kk}.  While this helps make the network topology easier
  to see in two-dimensional projections, these distances should not be
  taken too seriously.}.  As a comparison, we also show a visualization
based on the raw (unnormalized) interlock count of common members in
Fig.~\ref{network2}.  In Fig.~\ref{network108}, we depict the 108th House
using normalized interlock.


Some of the connections depicted in Figs.~\ref{network}--\ref{network108}
are unsurprising.  For instance, sets of subcommittees of the same standing
committee typically share many of the same members, thereby forming a group
or clique in the network.  The four subcommittees of the 107th Permanent
Select Committee on Intelligence, for example, each include at least half
of the full 20-member committee and at least one third of each of the other
subcommittees.  These tight connections result in normalized interlocks
with values in the range 14.4--23.6, causing these five nodes to be drawn
close together in the visualizations, forming the small pentagon in the
middle right of Fig.~\ref{network} and lower right of Fig.~\ref{network2}.
The Intelligence Committee and its subcommittees are also tightly connected
in the 108th House, appearing again as a small pentagon in
Fig.~\ref{network108}.

Some connections between committees, however, are less obvious.  For
instance, the 9-member Select Committee on Homeland Security, formed in
June 2002 during the 107th Congress in the aftermath of the terrorist
attacks of September 11, 2001~\cite{homeland}, has a strong connection to
the 13-member Rules Committee (with a normalized interlock of 7.4 from two
common members), which is the committee charged with deciding the rules and
order of business under which legislation is considered by other committees
and the full House~\cite{thomas}.  The Homeland Security Committee is also
connected to the 7-member Legislative and Budget Process Subcommittee of
Rules by the same two common members (with normalized interlock~13.7).  In
the 108th Congress (see Fig.~\ref{network108}), the Homeland Security
Committee swelled to 50 members but maintained a close association with the
Rules Committee (with a normalized interlock of 4.1 from 6 common members).

\section{The hierarchical structure of committees}
\label{structure}

We now turn to an examination of community structure in the networks of
committees based on the one-mode projection described above.  We do this
using several methods of hierarchical clustering, in which one begins with
a network and ends up (by construction) with a hierarchical (tree)
structure.  In this section, we discuss the hierarchical clustering method
known as \emph{single linkage clustering}~\cite{cluster}.  We found similar
results using several alternative community detection methods, which are
discussed in detail in Appendix~B.  For each of these methods, we quantify
the organizational structures we find using Horton-Strahler numbers (to
indicate the number of hierarchical levels) and modularity (to indicate the
compartmentalization into different groups).

\begin{figure}
\centerline{
\includegraphics[width=\textwidth]{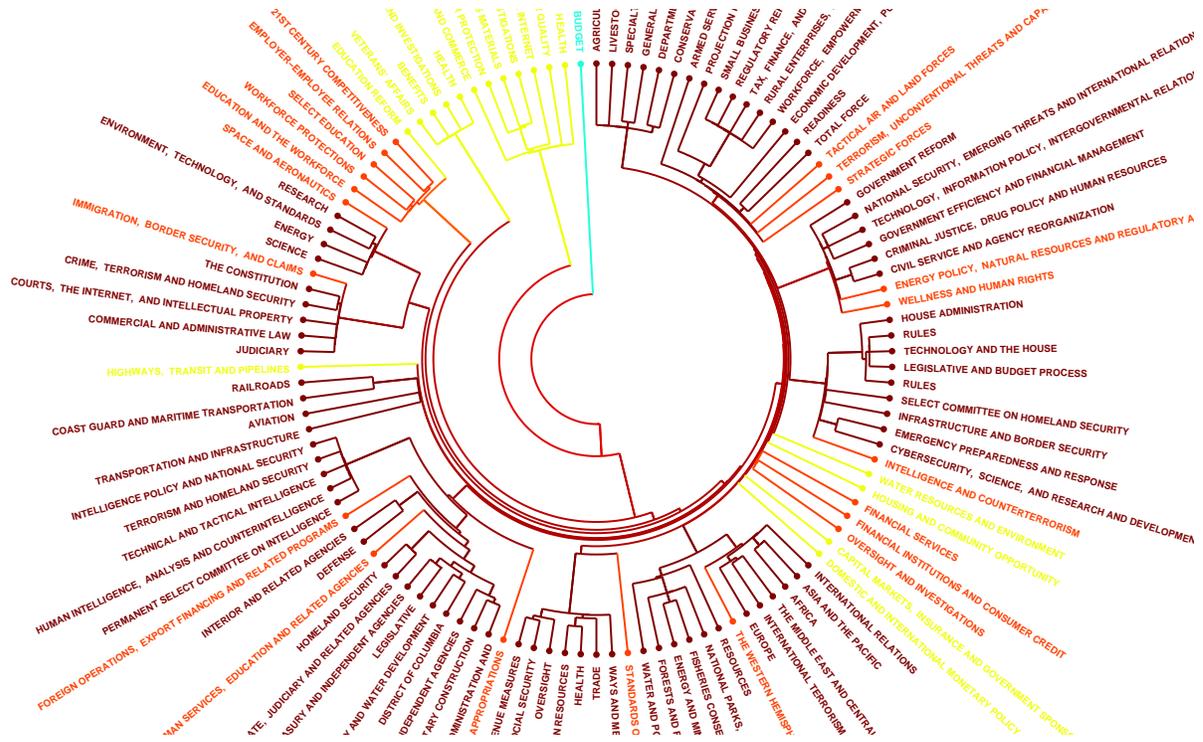}}
\caption{(Color) Dendrogram of the 108th House, determined by single
  linkage clustering and color-coded according to the Horton-Strahler
  values~\cite{horton,strahler} of its leaves (discussed in the text), with
  lower values in blue and higher values in red.
  The ties between Rules and Homeland Security persist (between the
  3 o'clock and 4 o'clock positions), despite the swelling in size of the
  latter committee to fifty members.  As discussed in Appendix~B, these
  ties are robust with respect to the algorithm used to determine the
  hierarchy.}
\label{108hs}
\end{figure}

To implement single linkage clustering, we start with the complete set of
committees for a given Congress.  We then join committees sequentially
starting with the pair with the greatest normalized interlock, followed by
the next greatest, and so forth.  This process generates ``clusters'' of
committees, which can be represented using a tree or \emph{dendrogram},
such as that shown in Fig.~\ref{107clust} for the 107th House.  Closer
examination of the dendrograms in Fig.~\ref{107clust} (from the 107th
Congress) and Fig.~\ref{108hs} (from the 108th Congress), each of which has a
Horton-Strahler number of 5, conveys four reasonably-expected and
well-ordered hierarchical levels of clustering: subcommittees, standing and
select committees, groups of standing and select committees, and the entire
House.  These single linkage clustering dendrograms also give some
suggestion of a weaker fifth level of organization corresponding to groups
of subcommittees inside larger standing committees.

In this paper, we are primarily interested in the organizational levels
describing the connections between standing and select committees (and
groups thereof).  For example, near the 8 o'clock position in
Fig.~\ref{107clust} (from the 107th Congress) is a tightly grouped cluster
that includes the House Rules Committee, the House Administration
Committee, and the Select Committee on Homeland Security.  A similar
cluster appears in Fig.~\ref{108hs} (from the 108th Congress) near the 3
o'clock position, including all five of the subcommittees of Homeland
Security introduced in the 108th Congress.  Because assignments to select
committees are ordinarily determined by drawing \emph{selectively} from
committees and subcommittees with overlapping jurisdiction, one might
naively expect a close connection between the Select Committee on Homeland
Security and, for example, the Terrorism and Homeland Security Subcommittee
of the Intelligence Select Committee, which was formed originally as a
bipartisan ``working group'' and was designated on September 13, 2001 by
Speaker of the House Dennis Hastert [R-IL] as the lead congressional entity
assigned to investigate the 9/11 terrorist attacks~\cite{homeland}.
However, the 107th Homeland Security Committee shares only one common
member (normalized interlock 2.4) with the Intelligence Select Committee
(located near the 1 o'clock position in Fig.~\ref{107clust}) and has no
interlock at all with any of the four Intelligence subcommittees.

As shown in Figure~\ref{107clust}, we can enrich the analysis by
color-coding each committee according to the mean ``extremism'' of its
members.  Extremism is determined using the results of a singular value
decomposition analysis of Representatives' voting records that places each
Representative on a scale that runs, roughly speaking, from the most
partisan Republican members of the House to the most partisan Democrats
(the SVD analysis is described in detail in Appendix~A).  The extremism of
a committee is then quantified as the average deviation of its members from
the mean on this scale.  Committees composed of highly partisan members of
either party appear in red in Fig.~\ref{107clust} and those containing more
moderate Representatives appear in blue.  Taking again the examples of
Intelligence and Homeland Security, we can immediately identify the former
as moderate and the latter as more partisan.  Indeed, the Select Committee
on Homeland Security has a larger mean extremism than any of the 19
standing Committees and has the 4th largest mean extremism among the 113
committees of the 107th House (see Table~\ref{xcom}).  This is perhaps not
so surprising when we see that its members included the House Majority
Leader, Richard Armey [R-TX], and both the Majority and Minority Whips, Tom
DeLay [R-TX] and Nancy Pelosi [D-CA].  However, our characterization of the
committee was made \emph{mathematically}, using no political knowledge
beyond the roll call votes of the 107th House.  As another example, the
107th House Rules Committee is the second most extreme of the 19 standing
committees (after Judiciary) and ranks 18th out of 113 committees overall.
In contrast, the Permanent Select Committee on Intelligence of the 107th
House has a smaller mean extremism than each of the 19 standing Committees,
and Intelligence and its four subcommittees all rank among the 10 least
extreme of all 113 committees.

\begin{table}
\centerline{
{\footnotesize
\begin{tabular}{|c|c|} \hline
 Most Extreme Committees  & Most/Least Extreme Committees and Subcommittees
\\ \hline
   Select Committee on Homeland Security & 1. Commercial and Administrative Law (Judiciary)  \\
   Judiciary & 2. Forests and Forest Health (Resources)  \\
   Rules & 3.Crime, Terrorism, and Homeland Security (Judiciary)   \\
   Standards of Official Conduct & 4. Select Committee on Homeland Security \\
   Resources &  5. Africa (International Relations) \\
   Budget & 6. Workforce Protections (Education and the Workforce)  \\
   Education and the Workforce & 7. Judiciary   \\
   Ways and Means & 8. Social Security (Ways and Means) \\
   International Relations & 9. Labor, Health and Human Services and Education (Appropriations) \\
   Small Business & 10. The Constitution (Judiciary)  \\
   House Administration &  \\
   Appropriations & 113. District of Columbia (Government Reform) \\
   Energy and Commerce & 112. Human Intelligence, Analysis and Counterintelligence (Intelligence) \\
   Financial Services & 111. Intelligence Policy and National Security (Intelligence) \\
   Government Reform & 110. Economic Development, Public Buildings and Emergency\\
   Armed Services & Management (Transportation and Infrastructure) \\
   Veterans' Affairs & 109. Technology and Procurement Policy (Government Reform) \\
   Science & 108. Technical and Tactical Intelligence (Intelligence) \\
   Transportation and Infrastructure & 107. Permanent Select Committee on Intelligence \\
   Agriculture & 106. General Farm Commodities and Risk Management (Agriculture) \\
   Permanent Select Committee on Intelligence & 105. Coast Guard and Maritime Transportation \\
   & (Transportation and Infrastructure) \\
   & 104. Terrorism and Homeland Security (Intelligence) \\ \hline
\end{tabular}}}
\caption{SVD rank ordering of the committees in the 107th House.  In the
first column, we list the standing and select committees from most extreme
to least extreme.  In the second column, we list the most extreme and least
extreme committees and subcommittees, with the parent committee shown in
parentheses when appropriate.  (The latter set of committees is listed from
less extreme to more extreme.)}
\label{xcom}
\end{table}

\section{Modularity}

To further investigate the observed hierarchies in the House committee
assignment networks, we employ the statistic known as~\emph{modularity},
modified to allow for the weighted nature of our networks.  Consider first
an unweighted network, which is divided into some number of groups of
vertices.  The modularity~$m$ for this division into groups is defined to
be~\cite{markfast}
\begin{equation}
  m = \sum_i (e_{ii} - a_i^2)\,,
\label{mod}
\end{equation}
where $e_{ij}$ denotes the fraction of ends of edges in group~$i$ for which
the other end of the edge lies in group~$j$ and $a_i = \sum_j e_{ij}$ is
the fraction of all ends of edges that lie in group~$i$.  Modularity
measures the difference between the total fraction of edges that fall
within---rather than between---groups (the first term) and the fraction one
would expect if edges were placed at random (respecting vertex degrees).
Thus, high values of the modularity indicate partitions of the network in
which more of the edges fall within groups than one would expect by chance.  This, in turn, has been found to be a good indicator of functional network
divisions in many cases~\cite{newmod}.

\begin{figure}
\begin{center}
\includegraphics[width = \textwidth]{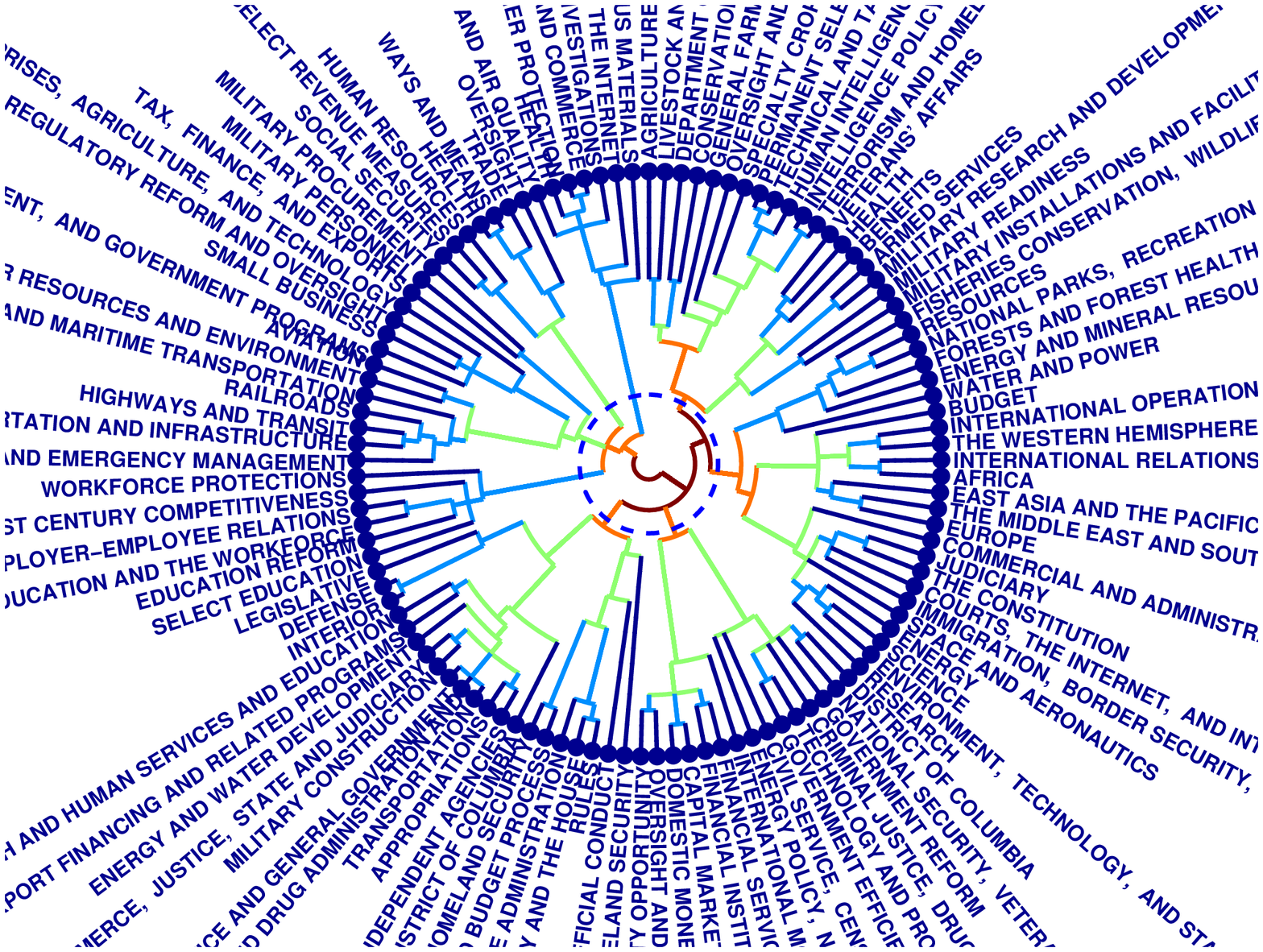}
\end{center}
\caption{(Color) Dendrogram of the committee assignment network for the
107th U.S.~House of Representatives, determined using the weighted
local community detection method (discussed in Appendix~B) with local clustering threshold $\alpha \approx 0.1179$. This value of $\alpha$ gives the dendrogram with highest
maximum modularity, as indicated by the dashed dividing ring.  The
dendrogram is color-coded according to the number of hierarchical
levels of each community in the tree, which is codified by
Horton-Strahler numbers (discussed in the text).  The Strahler numbers of
the communities are calculated as one moves from the outside to the
center of the tree.  When two nodes of Strahler number 1 (dark blue)
combine, they form a community with Strahler number 2 (light blue).
We also find communities with Strahler numbers of 3 (green), 4
(orange), and 5 (maroon), indicating the five hierarchical levels in the
committee assignment network of the 107th House.}
\label{modcolor}
\end{figure}

The projected one-mode networks we consider here are weighted.  In our
calculations, we therefore employ the weighted generalization of modularity described
in~\cite{newweight}, in which instead of counting numbers of edges falling
between particular groups, we count the sums of the weights of those edges,
so that heavily weighted edges contribute more than lightly weighted ones.
Both $e_{ij}$ and $a_i$ can be generalized in this fashion in a
straightforward manner, and then the modularity is again calculated from
Eq.~(\ref{mod}).  The meaning of the modularity remains essentially the
same: it measures when a particular division of the network has more edge
weight within groups than one would expect on the basis of chance.

We use modularity to quantify the organizational divisions of the networks
and to compare the dendrograms to each other.  In particular, the
modularity values shown in Table~\ref{modcompare} indicate that the
dendrograms produced via single linkage clustering have a better-defined
community structure (higher modularity) in the Republican-controlled Houses
(104th--108th) than in the Democrat-controlled ones.  Hence, with respect
to the metric of normalized interlock, the committee reorganization
following the Republican Revolution (which we already saw in
Fig.~\ref{revo}a produced a sharp decline in the typical numbers of
committee assignments per Representative compared to the 101st--103rd
Houses) seems also to have tightened the compartmentalization of the House
committee assignments.

\begin{table}
\centerline{
\begin{tabular}{|c|c|c|c|c|} \hline
Method: & SL & EB & NB & WL \\ \hline
$M_1\,,\,\tilde{M}_1$ & 0.1666\,, 0.1037 & 0.1450\,, 0.0904 & 0.1775\,, 0.0964 & 0.3388\,, 0.1945 \\ \hline
$M_2\,,\,\tilde{M}_2$ & 0.1831\,, 0.0923 & 0.1552\,, 0.0855 & 0.1693\,, 0.0979 & 0.3395\,, 0.1675 \\ \hline
$M_3\,,\,\tilde{M}_3$ & 0.1824\,, 0.0975 & 0.2640\,, 0.1388 & 0.2385\,, 0.1308 & 0.3884\,, 0.2343 \\ \hline
$M_4\,,\,\tilde{M}_4$ & 0.4221\,, 0.2077 & 0.2987\,, 0.1542 & 0.3312\,, 0.1909 & 0.4432\,, 0.2614 \\ \hline
$M_5\,,\,\tilde{M}_5$ & 0.3331\,, 0.1681 & 0.2518\,, 0.1290 & 0.2439\,, 0.1600 & 0.3945\,, 0.2228 \\ \hline
$M_6\,,\,\tilde{M}_6$ & 0.2982\,, 0.1709 & 0.2481\,, 0.1305 & 0.2420\,, 0.1581 & 0.3720\,, 0.1861 \\ \hline
$M_7\,,\,\tilde{M}_7$ & 0.3350\,, 0.1755 & 0.2604\,, 0.1293 & 0.2386\,, 0.1465 & 0.3748\,, 0.2133 \\ \hline
$M_8\,,\,\tilde{M}_8$ & 0.3324\,, 0.1736 & 0.2319\,, 0.1178 & 0.2294\,, 0.1417 & 0.3781\,, 0.2204 \\ \hline
\end{tabular}}
\caption{Maximum ($M_i$) and average ($\tilde{M}_i$) modularities of
  community structure for committee assignment networks for the 101st--108th
  House of Representatives, with dendrograms produced using single linkage
  clustering (SL), random-walk betweenness with sequential edge (committee
  assignment) removal (EB), random-walk betweenness with sequential
  Representative node removal (NB), and a generalization of the local
  community detection algorithm of~\cite{bagrow} to weighted networks (WL).
  These last three algorithms are described in Appendix~B.  For the WL
  method, we use the value of the local clustering threshold parameter
  $\alpha$ giving the greatest maximum modularity (see Appendix~B).}
\label{modcompare}
\end{table}

\section{Number of hierarchical levels}
\label{number}

Another interesting feature of dendrograms is the depth of their
hierarchical organization, which can be quantified by computing
Horton-Strahler numbers~\cite{horton,strahler,strahlernet,strahlernet2}.\footnote{The dendrograms we produce are perfectly hierarchical by construction, and it is accordingly important to provide a precise measurement of the depth of this organization.  As discussed in the main text, the levels we observe have natural interpretations based on the known hierarchical organization of the House of Representatives.}  As originally defined, Horton-Strahler numbers give the number of levels in
the minimum-depth branch of a tree.  Here we consider two generalizations.
First, we examine the Strahler numbers of leaves (see, for example,
Fig.~\ref{108hs}), assigning a value $S_j$ that identifies the number of
hierarchical levels associated with the $j$th subcommittee; this is the
number of levels in the particular branch of the tree with that specific
subcommittee as the leaf.  In Table~\ref{HStable} of Appendix~B, we compare
$S = \max_j S_j$, the mean $\tilde{S} = \langle S_j\rangle$, and the
standard deviation $\sigma = \langle(S_j-\tilde{S})^2\rangle^{1/2}$ in the
101st--108th Houses for single linkage clustering, two betweenness-based
dendrograms, and our local community detection method to quantify the
statistics of the hierarchical levels revealed by each method.  Second, we
also define a notion of Strahler numbers for communities (see
Fig.~\ref{modcolor}), in which a given subtree (i.e.,~community) is
assigned a Strahler number as if it were itself a full dendrogram.

As indicated previously, single linkage clustering identifies several
hierarchical levels of organization within the House of Representatives:
subcommittees, committees, and the floor (the whole House).  In all eight
Houses, we also identify a fourth hierarchical level, representing groups
of closely-connected committees.  In the 104th, 105th, 107th, and 108th
Houses, single linkage clustering reveals a fifth level of organization.
See, for example, Fig.~\ref{108hs}, which is color-coded according to the
$S_j$ values of the leaves/committees, and Fig.~\ref{modcolor}, which is
color-coded according to the Horton-Strahler $S$ values computed as if each
community were itself an individual tree.  Single linkage clustering
appears to organize the House's hierarchical structure more sharply than
betweenness-based dendrograms, as the trees produced by the former have
consistently higher values of $S$ and~$\tilde{S}$.  Additionally, networks
with high maximum and average Strahler numbers tend also to have high
modularity scores (compare Tables~\ref{modcompare} and~\ref{HStable}),
signifying their strong organizational structure.

Strahler numbers reveal additional information about the changes in the
House committee assignment networks after the Republican Revolution.  The
three lowest mean Strahler values occur in the Democrat-majority Houses
(101st--103rd), despite the fact that the number of committees and
subcommittees \emph{decreased} after the Republicans gained control of the
House (see Fig.~\ref{revo}).  In a perfectly balanced binary tree, one
would instead expect an increase in the Strahler number when nodes are
added.  Furthermore, all of the Republican-controlled Houses except the 106th have five levels of hierarchical structure (based on the metric of Strahler numbers) rather than the four revealed in the 101st--103rd Houses, so it seems that the change in majority party added an extra level
of hierarchical structure to the committee assignment network.

\section{Conclusions}
\label{conc}

We have applied methods from network theory, coupled with an SVD analysis
of roll call votes, to investigate the organizational structure of the
committees and subcommittees of the U.S.~House of Representatives.  Using
the 101st--108th Congresses as examples, we have found evidence of several
levels of hierarchy within the network of committees and---without
incorporating any knowledge of political events or positions---identified
some close connections between committees, such as that between the House
Rules Committee and the Select Committee on Homeland Security in the 107th
and 108th Congresses.  We have also identified correlations between
committee assignments and Representatives' political positions and examined
changes in the network structure across different Congresses, emphasizing
effects that resulted from the shift in majority control from Democrats to
Republicans starting with the 104th House.

\section*{Acknowledgements}
 
We thank Gordon Kingsley for challenging ideas that prompted some of this
research; Michael Abraham for developing some of the computer codes; Casey
Warmbrand for converting data into a usable format; Thomas Callaghan for
his computer codes and comments on this manuscript; Ron Burt, Aaron
Clauset, Sharad Goel, Debra Goldberg, Chris Wiggins, and Yan Zhang for useful
conversations; and James Fowler for a critical reading of this manuscript.
We thank two anonymous referees for critical comments that lead to changes
that have improved the paper.  We also acknowledge support provided by a
National Science Foundation VIGRE grant awarded to the School of Mathematics at
Georgia Tech.  MAP was also supported in part by the Gordon and Betty Moore
Foundation through Caltech's Center for the Physics of Information.  PJM
was also supported by start-up funds provided by the Institute for Advanced
Materials, Nanoscience and Technology and the Department of Mathematics at
the University of North Carolina at Chapel Hill.  We obtained the roll call
data for the 102nd--107th Congresses from the Voteview website
\cite{voteview}, the roll call data for the 101st Congress from the
Inter-University Consortium for Political and Social Research \cite{icpsr},
and the committee assignments for the 101st--108th Congresses from the web
site of the House of Representatives Office of the Clerk~\cite{clerk}.

\appendix
\section{Voting patterns}\label{vote}

In this appendix, we analyze the voting records of Representatives.  The
results of this analysis were used in the main text to investigate the
relationship between the network of interlocks linking the Congressional
committees and the political positions of their constituent
Representatives.

One way to characterize political positions is to tabulate individuals'
voting records on selected key issues (via, for example, interest group
ratings), but such a method is subjective by nature and a procedure that
involves less personal judgment is preferable.  Here we apply the
``multi-dimensional scaling'' technique known as singular value
decomposition (SVD)~\cite{vanloan,sirovich} to the complete voting records
of each session of the House~\cite{pr,pr00}.  Each two-year term of
Congress is treated in isolation from the others.  Other methods of
analysis~\cite{senatemine,finnmine}, such as the Bayesian approach of
\cite{jackman}, also yield useful results.  The advantages of
multi-dimensional scaling techniques versus factor analysis (which has a
long tradition in political science) in analyzing voting data are discussed
in detail in~\cite{brazill}.

We define an $n\times m$ voting matrix $\mathbf{B}$ with one row for each
of the $n$ Representatives in the House and one column for each of the $m$
votes taken during a two-year term.  For instance, the 107th House had
$n=444$ Representatives (including midterm replacements) and took $m=990$
roll call votes.  The element $B_{ij}$ is $+1$ if Representative~$i$ voted
``yea'' on measure~$j$ and $-1$ if he or she voted ``nay.''  If a
Representative did not vote because of absence or abstention, the
corresponding element is~$0$.  (We do not separately identify abstentions
from absences; additionally, a relatively low number of false zeroes is
generated by resignations and midterm replacements.)

The SVD identifies groups of Representatives who voted in a similar fashion
on many measures.  The grouping that has the largest mean-square overlap
with the actual groups voting for or against each measure is given by the
leading (normalized) eigenvector $\mathbf{u}^{(1)}$ of the
matrix~$\mathbf{B}^T\mathbf{B}$, the next largest by the second eigenvector
$\mathbf{u}^{(2)}$, and so on~\cite{vanloan,sirovich}.  If we denote by
$\sigma_k^2$ the corresponding eigenvalues (which are provably
non-negative) and by $\mathbf{v}^{(k)}$ the normalized eigenvectors of
$\mathbf{B}\mathbf{B}^T$ (which have the same eigenvalues), then it can be
shown that
\begin{equation}
	B_{ij} = \sum_{k=1}^n \sigma_k u^{(k)}_i v^{(k)}_j\,,
\label{decomposition}
\end{equation}
where $\sigma_k\ge0$ for all~$k$.  The matrix~$\mathbf{B}^{(r)}$, $r<n$,
with elements
\begin{equation}
	B^{(r)}_{ij} = \sum_{k=1}^r \sigma_k u^{(k)}_i v^{(k)}_j \label{full}
\end{equation}
approximates the full voting matrix~$\mathbf{B}$.  The sum of the squares
of the errors in the elements is equal to $\sum_{k=r+1}^n \sigma_k^2$,
which vanishes in the limit $r \to n$.  Assuming the quantities~$\sigma_k$,
called the \textit{singular values}, are ordered such that
$\sigma_1\ge\sigma_2\ge\sigma_3\,\ldots$, this implies that
$\mathbf{B}^{(r)}$ is a good approximation to the original voting matrix
provided the singular values decay sufficiently rapidly with
increasing~$k$.  Alternatively, one can say that the $l$th term in the
singular value decomposition~(\ref{decomposition}) accounts for a fraction
$\sigma_l^2/\sum_{k=1}^n \sigma_k^2$ of the sum of the squares of the
elements in the voting matrix.

\begin{figure}
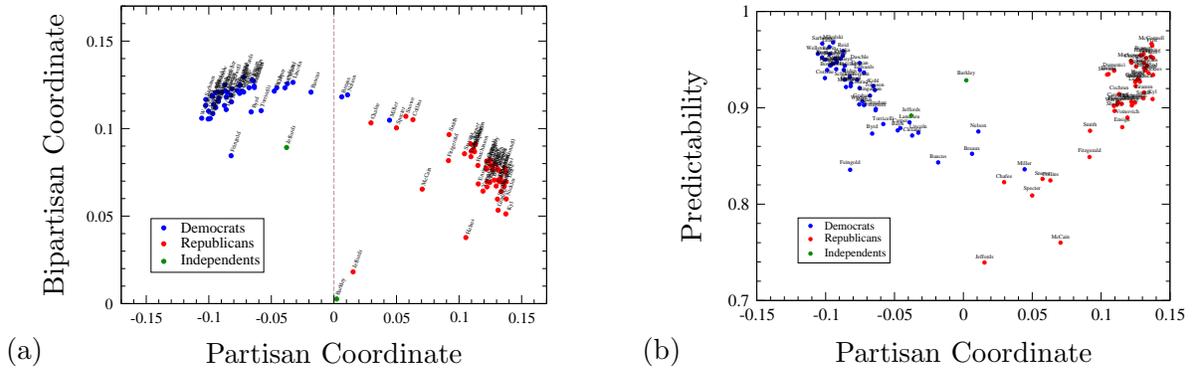

  {\psfragscanon
  \psfrag{107th Senate}{\hspace*{1.2in}107th Senate}
  \psfrag{partisan}[t][t]{Partisan Coordinate}
  \psfrag{partisan coordinate}[t][t]{Partisan Coordinate}
  \psfrag{bipartisan}[b][b]{Bipartisan Coordinate}
  \psfrag{predictability}[b][b]{Predictability}
  \centerline{(a)
\includegraphics[width=0.4\textwidth]{senate107labeled.eps} \hspace{1cm}
  (b) \includegraphics[width=0.4\textwidth]{predict107.eps}}}
\caption{(Color) Singular value decomposition (SVD) of the Senate voting
  record from the 107th U.S.~Congress.  (a) Two-dimensional projection of
  the voting matrix.  Each point represents a projection of a single
  Representative's votes onto the leading two eigenvectors (labeled
  ``partisan'' and ``bipartisan,'' as explained in the text).  Democrats
  (blue) appear on the left and Republicans (red) are on the right.
  (Independents are shown in green).  (b) ``Predictability" of votes cast
  by Senators in the 107th Congress based on a two-dimensional projection
  of the SVD.  Individual Senators range from 74\% predictable to 97\%
  predictable.}\label{senate}
\end{figure}

To an excellent approximation, we find that a Representative's voting
record can be characterized by just two coordinates.  That is,
$B^{(2)}_{ij}$~is a good approximation to~$B_{ij}$.
Observing that one of the two directions correlates well with party
affiliation for members of the two major parties, we call this the
``partisan'' coordinate.  We call the other direction the ``bipartisan''
coordinate, as it correlates well with how often a Representative votes
with the majority.  Because Senators are generally better known than
Representatives, we plot as an example the coordinates along these first
two eigenvectors for the 107th Senate in Fig.~\ref{senate}a.  As expected,
Democrats are grouped together and are almost completely separated from
Republicans.  For ease of identification, we follow the sign convention
that places Democrats on the left and Republicans on the right.  The few
instances of apparent party misidentification by the partisan coordinate in
Fig.~\ref{senate} are unsurprising.  Conservative Democrats, such as Zell
Miller [D-GA], appear farther to the right than some moderate
Republicans~\cite{org}.  Additionally, Senator James Jeffords [I-VT], who
left the Republican party to become an Independent early in the 107th
Congress, appears closer to the Democratic group than the Republican one
and to the left of several of the more conservative Democrats.  (He appears
twice in Fig.~\ref{senate}a, once each for votes cast under his two
different affiliations.)

One can use a truncation of the SVD to construct an approximation to the
votes in the full roll call~\cite{sirovich}.  For instance, with our
two-dimensional approximation to the voting matrix, we assign ``yea''
or ``nay'' votes to individuals based on the signs of the corresponding
elements of the matrix.  In Fig.~\ref{senate}b, we show the fraction of
actual votes correctly reconstructed by this approximation, which gives a
measure of the ``predictability'' of the Senators in the 107th Congress.
For both parties, moderate Senators are less predictable than hard-liners.
The two-dimensional projection correctly reconstructs the votes of some
hard-line Senators for as many as 97\% of the votes they cast.  Examining
the apparent outliers in Fig.~\ref{senate}b, the votes Senator Jeffords
cast as a Republican appear here to make him the least ``predictable''
Senator.  However, it is important to emphasize that Jeffords cast
relatively few votes as a member of the Republican party, so it is not
surprising that this behavior is less predictable because the voting record
includes a large number of artificial absences.  The other Independent in
Fig.~\ref{senate}b is Senator Dean Barkley [I-MN], who completed the
remainder of the term for Senator Paul Wellstone [D-MN] in the 107th
Congress after Wellstone's death.  While one might be tempted to interpret
Barkley's partisan coordinate as balanced, its value is strongly influenced
by the large number of effective absences in the SVD analysis because he
was not appointed until very late in the term.  Both of his first two
coordinates consequently lie near zero.  The other apparent outliers in
Fig.~\ref{senate}b---Senators Russ Feingold [D-WI] and John McCain
[R-AZ]---are both known for their occasional ``maverick'' behavior.

\begin{figure}
{\psfragscanon
  \psfrag{Partisan Coordinate}[t][t]{Partisan Coordinate} 
  \psfrag{Bipartisan Coordinate}[b][b]{Bipartisan Coordinate}
  \centerline{
  \includegraphics[width = .9\textwidth]{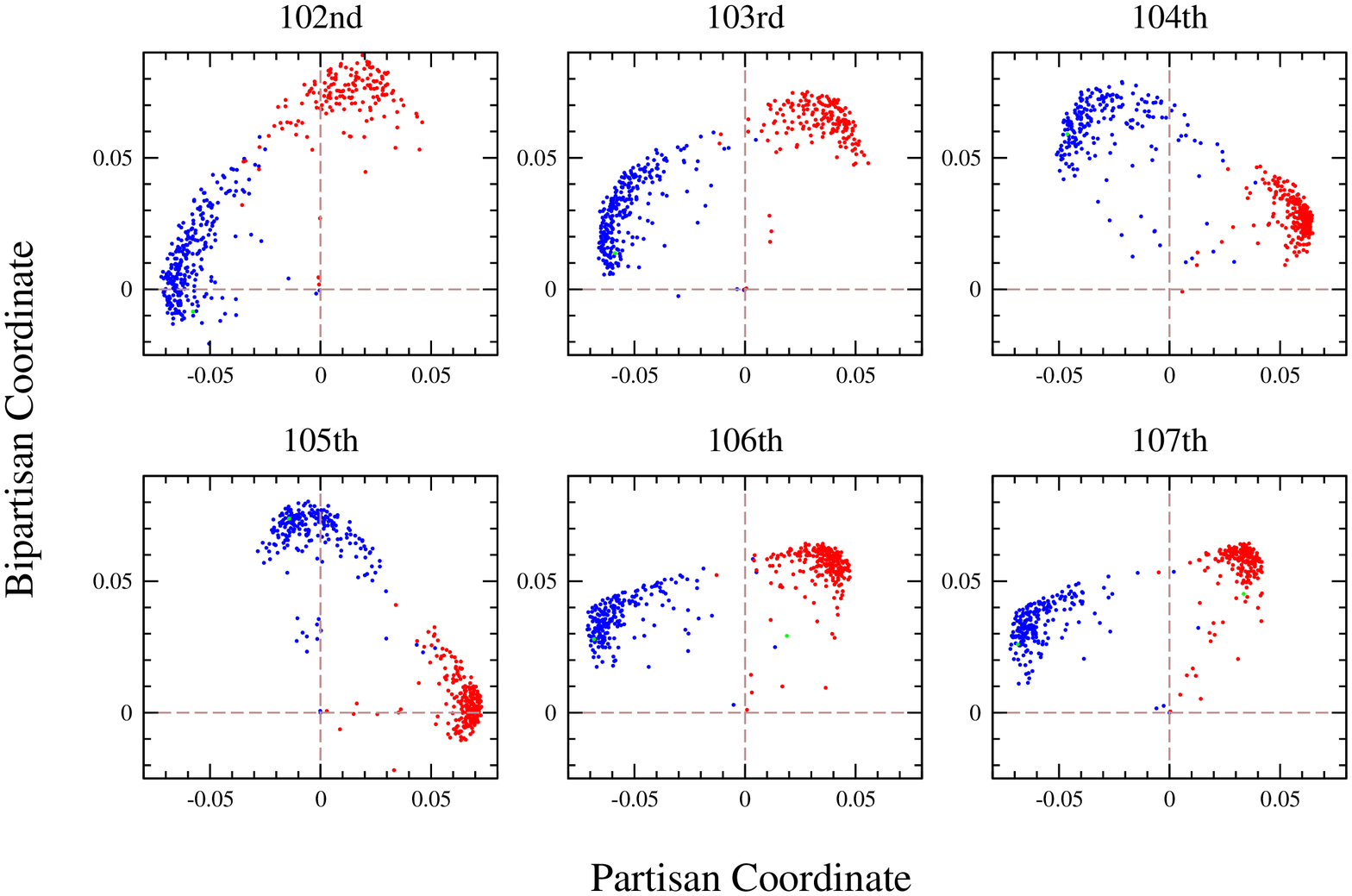}}}
\caption{(Color) SVD of the voting record for the House of Representatives for each of the 102nd--107th U.S.~Congresses.
  As with the Senate, Democrats (blue) appear on the left and
  Republicans (red) are on the right (with Independents shown in green).}
\label{svdeg}
\end{figure}

Having demonstrated the application of SVD to the analysis of the voting
records of the Senate, let us now return to the House of Representatives.
For the 107th House, we find that the leading eigenvector accounts for about 45.3\% of the variance of the voting matrix, the second eigenvector accounts for about 29.6\%, and no other
eigenvector accounts for more than 1.6\%.  We obtain similar results for
other recent Congresses, with two eigenvectors giving a good approximation
to the voting matrix in every case (see Table~\ref{per}).  In
Fig.~\ref{svdeg}, we plot these two coordinates for every member of the
House of Representatives for each of the 102nd--107th Congresses.  It has
been shown previously using other methods that Congressional voting
positions are well-approximated by just two coordinates~\cite{pr,voteview},
but it is important to emphasize that different identification methods
treat the ``bipartisan'' direction differently.  In particular, some
methods eliminate it entirely and associate the two remaining dimensions
with partisanship and an additional direction often identified as a
North-South axis, which was historically important during periods of
heightened concern about civil rights~\cite{pr,voteview}.  The SVD analysis
here keeps the ``bipartisan'' coordinate, making identifications in a
particularly simple fashion straight from the voting matrix containing the
roll call data.

\begin{table}
\centerline{
{\footnotesize
\begin{tabular}{|c|c|c|c|c|c|c|} \hline
Congress & 1st & 2nd & 3rd & 4th & 5th & 6th \\ \hline
101st & 40.0 & 20.4 & 3.0 & 1.7 & 1.5 & 1.1 \\ \hline
102nd & 39.6 & 20.1 & 2.8 & 1.8 & 1.2 & 1.0 \\ \hline
103rd & 43.1 & 21.5 & 2.9 & 1.6 & 1.3 & 0.9 \\ \hline
104th & 47.1 & 19.1 & 2.7 & 1.6 & 1.3 & 0.7 \\ \hline
105th & 38.6 & 28.7 & 2.0 & 1.5 & 1.2 & 0.9 \\ \hline
106th & 40.0 & 29.5 & 1.9 & 1.6 & 1.2 & 0.9 \\ \hline
107th & 45.3 & 29.6 & 1.6 & 1.4 & 1.0 & 0.7 \\ \hline
\end{tabular}}}
\caption{Amount of voting information, measured as the percentage of the
variance of the voting matrix, encoded by the six leading eigenvectors of
the 101st--107th House voting matrices.  The first column indicates the
Congress, and the next six columns give the percentage of information
encoded by each of the six leading eigenvectors.} \label{per}
\end{table}

As with the Senate, we find that the leading eigenvector corresponds closely to the
acknowledged political party affiliation of the Representatives, with
Democrats (blue) on the left and Republicans (red) on the right in the
plots~\footnote{This holds for the 101st--105th Houses.  The leading and
second eigenvectors switch roles in the 106th and 107th Houses.}.
Representatives who score highly on this ``partisan" coordinate---either
positively or negatively---tend often to vote with members of their own
party.  From this coordinate, we also compute a measure of ``extremism'' for
each Representative as the absolute value of their partisan coordinate
relative to the mean partisan score of the full House.  That is, we define
the extremism $e_i$ of a Representative by $e_i = |p_i - \mu|$, where $p_i$
is the Representative's partisanship score and $\mu$ is the mean value
(usually skewed slightly towards the majority party) of that coordinate for
the entire House.  In Table~\ref{part}, we list the most and least partisan
Representatives from each party (as computed from the roll call) for the
107th House.

By contrast, the second eigenvector groups essentially all Representatives
together regardless of party affiliation and thus appears to represent
voting actions in which most members of the House either approve or
disapprove of a motion simultaneously.  Representatives who score highly on this ``bipartisan coordinate" tend to often vote with the majority of the House.

The mean extremism for the Representatives in the 107th House is about
$0.0458$, and the standard deviation is $\sigma \simeq 0.0090$.  The
extremism of committees as averages over their constituent members yields a
distribution of mean $0.0456$ and standard deviation $0.0032$.  By
contrast, one might crudely expect the standard deviation of the
committees' extremism to be approximately $\sigma/\sqrt{N}$, where $N$ is
the average number of Representatives per committee.  For the 107th House,
this gives $0.0090/\sqrt{21.87} \simeq 0.0019$.  Hence, the distribution of
committee extremism is roughly 50\% wider than what would be expected with
independent committee assignments.  Basic statistics concerning committee
extremism for the 101st--107th Houses are summarized in
Table~\ref{extreme}.  Observe, for example, that the relative variance
versus that expected from random committee and subcommittee assignments
increases with every Congress (with the largest increase occurring between
the 106th and 107th Houses).

\begin{table}
\centerline{
{\footnotesize
\begin{tabular}{|c|c|c|} \hline
Least Partisan & Farthest Left & Farthest Right \\ \hline
K. Lucas [R] & J.~D. Schakowsky [D] & T.~G. Tancredo [R] \\ 
C.~A. Morella [R] & J.~P. McGovern [D]  & J.~B. Shadegg [R]\\ 
R.~M. Hall [D] & H.~L. Solis [D] & J. Ryun [R] \\ 
R. Shows [D] & L.~C. Woolsey [D] & B. Schaffer [R] \\ 
G. Taylor [R] & J.~F. Tierney [D] & P. Sessions [R]  \\ 
C.~W. Stenholm [D] & S. Farr [D] & S. Johnson [R]  \\ 
R.~E. Cramer [D]  & N. Pelosi [D]   & B.~D. Kerns [R]  \\ 
V.~H. Goode [R] & E.~J. Markey [D]  & P.~M. Crane [R] \\ 
C. John [D] & J.~W. Olver [D] & W.~T. Akin [R] \\ 
C.~C. Peterson [D] & L. Roybal-Allard [D] & J.~D. Hayworth [R] \\ \hline
\end{tabular}}}
\caption{SVD rank ordering of the most and least partisan Representatives in
the 107th U.S.~House.  The first column gives the least partisan
Representatives, as determined by an SVD of the roll call votes.  The second
column gives the SVD rank ordering of the most partisan Representatives.
They are all Democrats (the mean partisanship is skewed slightly towards the Republican party because it held the majority), so this also ranks the Representatives
farthest to the Left.  The third column gives the rank ordering of the Representatives
farthest to the Right.}  \label{part}
\end{table}

\begin{table}
\centerline{
{\footnotesize
\begin{tabular}{|c|c|c|c|c|c|c|c|} \hline
House & $\mu_R$\,, Var$_R$ & $\mu_C$\,, Var$_C$ & $\mu_S$\,, Var$_S$ &
$\langle C\rangle$ & $\Delta$(Var$_C$) & $\langle S\rangle$ & $\Delta$(Var$_S$) \\ \hline
   101st & 0.0252\,, 0.000167 & 0.0248\,, $1.458 \times 10^{-5}$  & 0.0251\,,
$1.454 \times 10^{-5}$ & 16.755 & 0.984 & 27.091 & 1.357  \\ \hline
   102nd & 0.0335\,, 0.000210 & 0.0335\,, $1.540 \times 10^{-5}$ & 0.0343\,,
$9.818 \times 10^{-6}$ & 17.632 & 0.856 & 37.773 & 1.403 \\ \hline
   103rd & 0.0423\,, 0.000194  & 0.0425\,, $1.544 \times 10^{-5}$ & 0.0429\,,
$9.014 \times 10^{-6}$ & 18.206 & 0.959 & 39.091 & 1.410 \\ \hline
   104th & 0.0431\,, 0.000171 & 0.0430\,, $1.262 \times 10^{-5}$  & 0.0435\,,
$7.830 \times 10 ^{-6}$ & 19.774 & 1.080 & 40.842 & 1.397  \\ \hline
   105th & 0.0340\,, 0.000108 & 0.0339\,, $8.927 \times 10^{-6}$ & 0.0345\,,
$6.059 \times 10^{-6}$ & 20.056 & 1.169 & 41.526 & 1.651 \\ \hline
   106th & 0.0455\,, 0.000128 & 0.0455\,, $1.298 \times 10^{-5}$ & 0.0462\,,
$5.686 \times 10^{-6}$ & 20.944 & 1.472 & 42.474 & 1.330 \\ \hline
   107th & 0.0458\,, $8.080 \times 10^{-5}$ & 0.0456\,, $1.008 \times 10^{-5}$ &
 0.0461\,, $4.358 \times 10^{-6}$ & 21.867 & 1.927 & 43.368 & 1.625  \\ \hline
\end{tabular}}}
\caption{Committee extremism statistics for the 101st--107th Houses.
The first column indicates the Congress.  The second gives the mean and
variance of the extremism of the Representatives in that Congress.  The
third and fourth columns give the mean and variance under the independence
assumption of, respectively, all the committees and only the standing
committees (without select committees).  The fifth column gives the average
committee size, and the sixth gives how much larger the variance of
committee extremism is than would be expected under an independence
assumption (as reported in column three).  That is, the variance of the
committees' extremism is this factor multiplied by the variance expected if
Representatives were assigned to committees independently at random.  The
seventh and eighth columns repeat these numbers for standing committees
(compare with column four).}  \label{extreme}
\end{table}

\begin{figure}
{\psfragscanon
  \psfrag{partisan coordinate}[t][t]{Partisan Coordinate}
  \psfrag{bipartisan coordinate}[b][b]{Bipartisan Coordinate}
\centerline{
  \includegraphics[width=.9\textwidth]{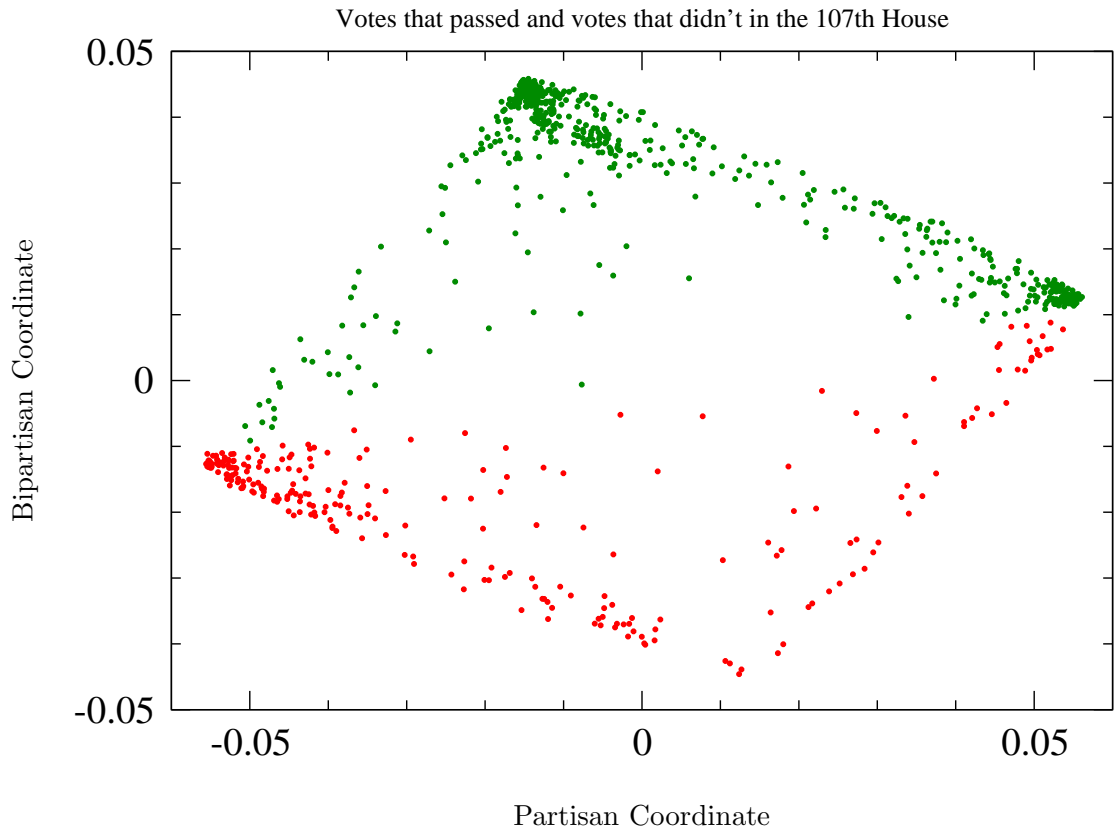}}}
\caption{(Color) SVD of the roll call of the 107th House projected onto the
  voting coordinates.  Points represent projections of the votes cast on a
  measure onto eigenvectors associated with the leading two singular
  values. There is a clear separation between measures that passed (green)
  and those that did not (red).  The four corners of the plot are
  interpreted as follows: measures with broad bipartisan support (north)
  all passed; those supported mostly by the Right (east) passed because the
  Republicans constituted the majority party of the 107th House; measures
  supported by the Left (west) failed because of the Democratic minority;
  and the (obviously) very few measures supported by almost nobody (south)
  also failed.}
\label{votesvd}
\end{figure}

Using the SVD results, we can also calculate the positions of the
\emph{votes} (as opposed to the voters) along the same two leading
dimensions to quantify the nature of the issues being decided.  We show
this projection for the 107th House in Fig.~\ref{votesvd}.  One application
of this analysis is a measurement of the reproducibility of individual
votes and outcomes.  Reconstituting the voting matrix as before using only
the information contained in the two leading singular values and the
corresponding eigenvectors and summing the resulting approximated votes
over all Representatives, we derive a single score for each vote.  Making a
simple assignment of ``pass'' to those votes that have a positive score and
``fail'' to all others successfully reconstructs the outcome of $984$ of
the $990$ total votes (i.e., about 99.4\% of them).  Overall, $735$ (about
74.2\%) of the votes passed, so simply guessing that every vote passed
would be considerably less effective.  Ignoring values from known absences
and abstentions (i.e., zeroes in the original voting matrix), the analysis
still identifies $975$ of the $990$ results correctly.  Even with the most
conservative measure of the reconstruction success rate---in which we
ignore values associated with abstentions and absences, assign
\emph{individual} yeas or nays according to the sign of the matrix
elements, and then observe which outcome has a majority in the resulting
roll call---the two-dimensional reconstruction still identifies $939$
(about 94.8\%) of the outcomes correctly.  We repeated these calculations
for the 101st--106th Houses and found similar results in each case (see
Table~\ref{recon}).  The remarkable accuracy of SVDs in reconstructing
votes was previously observed for U.S.~Supreme Court cases
in~\cite{sirovich}.  The Optimal Classification (OC) technique
of~\cite{pr00} (see also \cite{voteview}) also generates a rank ordering of the Representatives in the 107th House and correctly classifies 92.8\% of the individual Representatives' votes.


\begin{table}
\centerline{
{\footnotesize
\begin{tabular}{|c|c|c|c|} \hline
Congress & Number of votes & Outcomes correctly reconstructed & \% Individual
votes reconstructed \\ \hline
101st & 879 & 867, 864, 848 & 87.6 \% \\ \hline
102nd & 901  & 892, 888, 850 & 87.6 \% \\ \hline
103rd & 1094 & 1075, 1072, 1021 & 88.9 \% \\ \hline
104th & 1321 & 1307, 1312, 1225 &  89.3 \% \\ \hline
105th & 1166 & 1157, 1164, 1079 &   89.7 \% \\ \hline
106th & 1209 & 1200, 1198, 1121 &   90.6 \% \\ \hline
107th & 990 & 984, 975, 939  & 92.7 \%  \\ \hline
\end{tabular}}}
\caption{Roll call outcome reconstruction in the 101st--107th House
of Representatives using two-coordinate projections of SVDs.  The first
column gives the Congress, and the second indicates the total number of
measures in their roll call.  In the third column, we show consecutively
the number of outcomes correctly identified from the unmodified
reconstruction, the number correctly identified throwing out known absences
and abstentions, and the number correctly identified throwing out absences
and abstentions and then reconstructing individual Representatives' yeas/nays
and taking a majority vote.  In the fourth column, we show the percentage
of individual votes correctly reconstructed (which we note increases slightly from one Congress to the next during this time period).} \label{recon}
\end{table}

In making the connection between the voting record and committee assignment
networks, we remark that we constructed the committee assignment networks
representing the 101st--107th Houses from documents obtained from the
website of the U.~S. House of Representatives Office of the
Clerk~\cite{clerk}, which were based on the committee assignments at the
end of each Congress.  The roll calls, by contrast, include votes from
Representatives who subsequently died or resigned and hence were not
present at the end of the session.  Our networks also include
Representatives (such as non-voting Delegates) who do not appear in the
voting record.  To combine the network structures with the political
spectra (as determined using the SVD analysis), it was thus necessary to reconcile the two data sets by removing a few Representatives in each of these categories (about 5--10 from each roll
call and 5--10 from each network).  In situations where we have incorporated political
spectra into our network analysis, it is always with this slightly abridged
set of Representatives.  The subsequent SVD computations show
little change as a result of these adjustments.

\section{Community detection algorithms}\label{detectapp}

The results in the main text of this paper use single linkage clustering to
determine the community structure of the network of committees, but
several other methods can also be used (see, for
example, \cite{structpnas,commreview} and references therein).
It is of interest to ask whether our results are robust with respect to changes in
the method employed.  To address this question, we have explored three
other methods: two based on ``betweenness'' measures calculated on the full
bipartite networks of Representatives and committees and a local community
detection algorithm for weighted networks, generalized from the method for
unweighted networks introduced in~\cite{bagrow}.  As we now describe, we
obtain similar groupings with these different algorithms, although with
some differences, suggesting that the large-scale features (but perhaps not
the details) observed in our single-linkage clustering calculations are
fundamental properties of the networks and not a result of our choice of
methodology.

\subsection{Betweenness-based community detection}\label{between}

Communities can be detected in many cases using ``betweenness'' measures
that iteratively pick out and remove high-traffic edges (or other network
components) that lie on a large number of paths between vertices.  Repeated
application of such a procedure eventually fragments a network into
components, with the entire process represented by dendrograms similar to
those generated by standard hierarchical
clustering~\cite{structpnas,structmix,structeval}.

We have performed a corresponding calculation modified slightly to respect
the bipartite nature of the committee assignment networks, for which
betweenness can be computed by counting the number of shortest paths
between pairs of committees that traverse each edge in the network.
Additionally, we compute betweenness from densities of random walks between
committees rather than from geodesics (see \cite{walkbetween}), in part
because the small diameter of the network often leads to many non-unique
geodesics.  We use this betweenness measure in two different algorithms.
In one, we sequentially remove those edges (i.e.,~committee assignments)
with highest betweenness.  In the other, we sequentially remove the nodes (i.e.,~Representatives) lying on the largest number of paths.  Applying these two methods to the full (unweighted) bipartite committee
assignment graphs avoids altogether the projection onto a one-mode network
and the definition of the normalized interlock used in single linkage
clustering.

\begin{figure}
\begin{center}
  \includegraphics[width=\textwidth]{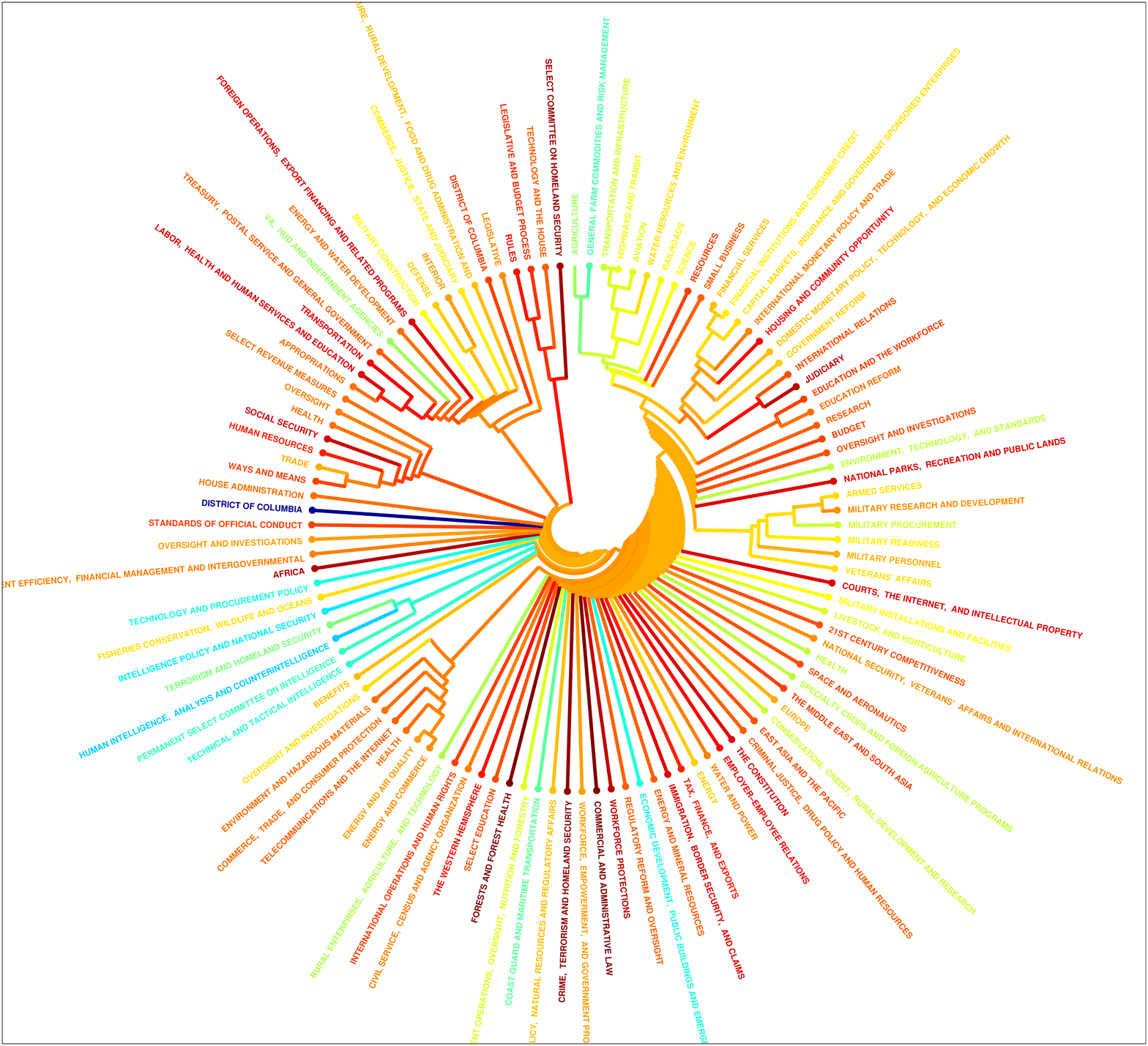}
\end{center}
\caption{(Color) Dendrogram of the committees of the 107th U.S.~House of
  Representatives constructed by sequentially removing individual
  committee assignments with highest random-walk betweenness (and
  subsequently recomputing betweenness).  Committees are listed
  counterclockwise around the outside of the figure in the order in
  which the algorithm separates them from the largest network
  component.  Committees and groups of committees are again
  color-coded according to their mean extremisms.  The first group of
  separated committees (just to the left of the 12 o'clock position)
  includes the Rules Committee and Select Committee on Homeland
  Security; this algorithm again indicates their close connection.}
\label{107biedge}
\end{figure}

Comparing the different community detection schemes, we see that the
dendrogram for the 107th House determined from random-walk betweenness and
edge removal (see Fig.~\ref{107biedge}) shows four levels of hierarchical
organization and again reveals the tight connections between the Rules
Committee (and its subcommittees) and the Select Committee on Homeland
Security.  We also again observe the close ties between the Intelligence
Committee and its subcommittees.  However, other connections seemingly
apparent in the single linkage clustering dendrogram are not uncovered by
this betweenness-based method.  Some subcommittees are not even grouped
with their parent committee; for example, near the 6 o'clock and 7 o'clock
positions, we see a weakly grouped cluster of committees that includes
(consecutively) the Forests and Forest Health Subcommittee of the Resources
Committee, the Select Education Subcommittee of the Education and the
Workforce Committee, and the Western Hemisphere Subcommittee of the
International Relations Committee.

Strahler numbers provide a way to quantify the robustness of some of these
different groupings.  The dendrogram representing the community structure
of the 107th House determined by random-walk edge removal has a Strahler
number $S = 4$.  Its average Strahler number of $\tilde{S} \simeq 2.7788$
quantifies the fact that many committees break off as singletons.  However,
the portion with the Rules Committee and Select Committee on Homeland
Security has a value of~3.  This grouping therefore gives meaningful
organizational information (in that it refers to an actual clique in the
network), even though the tree as a whole does not show a tremendous amount
of hierarchical structure.

\subsection{Weighted Local Community Detection}\label{other}

We have also constructed dendrograms from the one-mode committee networks
using a local community-detection algorithm generalized from a method for
unweighted networks developed by Bagrow and Bollt~\cite{bagrow}.  The goal
of this algorithm is to find a highly connected set of nodes (a ``local
community'') near each node and to combine these individual (potentially
overlapping) communities for each node into a hierarchical community
structure.  We again use the network of committees weighted by normalized
interlocks that we considered for single linkage clustering.

To detect communities, we start with a given House's (one-mode) adjacency
matrix~$\mathbf{A}$, whose element $A_{ij}$ gives the normalized interlock
between the $i$th and $j$th committees.  For convenience, we further
normalize these elements by the maximum normalized interlock, so that
$0\leq A_{ij}\leq 1$.  We use these weights as inverse distances to compute
a distance matrix~$\mathbf{D}$, where the element $D_{ij}$ designates the
shortest distance along any path from the $i$th node to the $j$th node.  We
then define a clustering coefficient $k$ of a selected group of $n$ nodes
as the sum of all weights within that group divided by $\frac{1}{2}
n(n-1)$.  In our generalization of the algorithm in \cite{bagrow}, we
define the \emph{$d$-shell} of node $i$ to be all nodes within distance $d$
of~$i$ according to the distance matrix.  We identify the local community
of the $i$th node to be the largest $d$-shell of node $i$ with $k \geq
\alpha$ for some threshold~$\alpha$.  As $\alpha$ is increased, the
definition of a local community becomes more stringent and smaller cliques
are obtained.








\begin{figure}
  \centerline{(a) \includegraphics[width=0.4\textwidth]{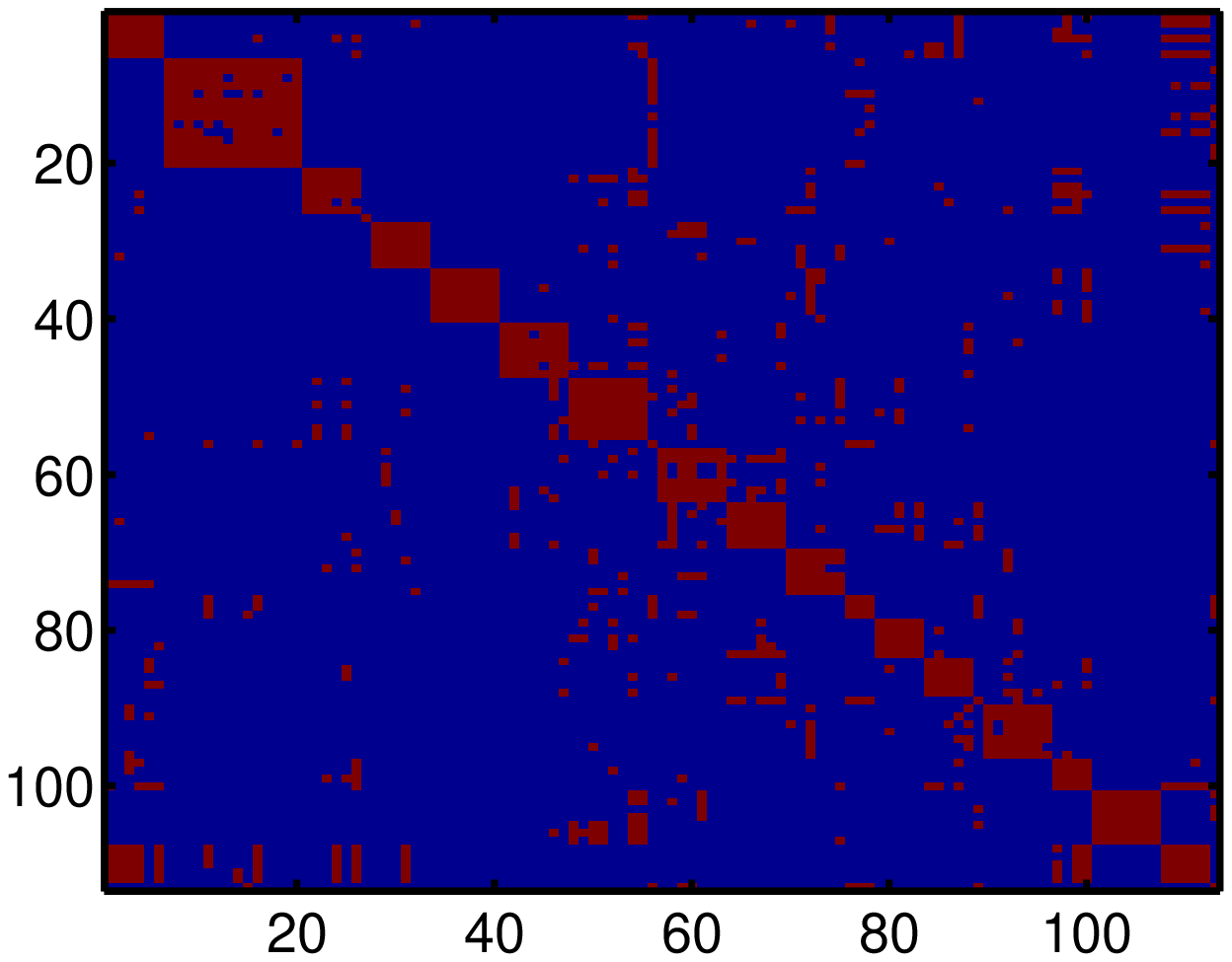} 
  \hspace{1cm}
  (b) \includegraphics[width=0.4\textwidth]{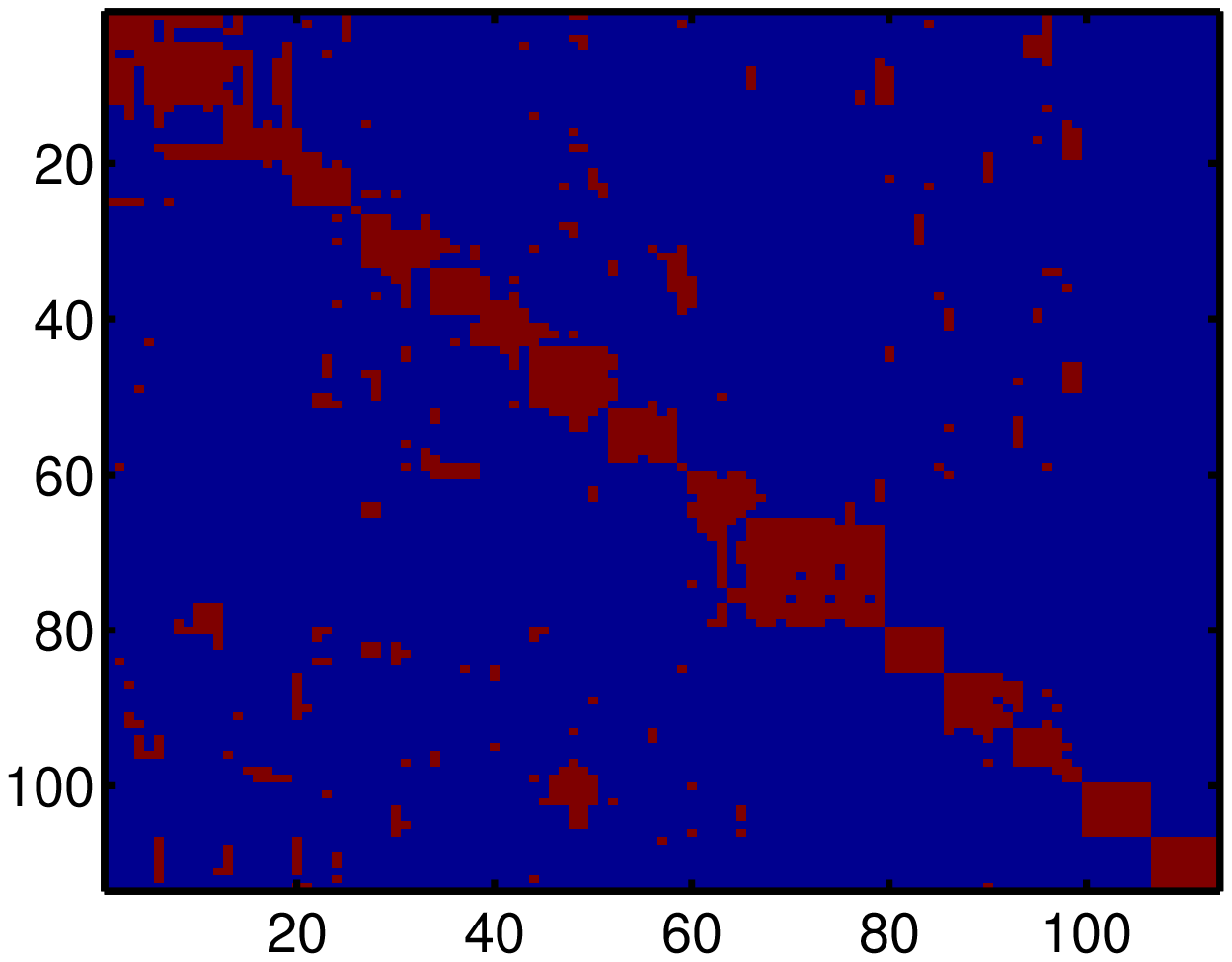}}
\caption{(Color) Visualization of the unsorted (a) and sorted (b) membership matrices for the 107th House.  The colors indicate the nearly full blocks of value 1 along the diagonal and the mostly zero-valued entries that are near the diagonal but outside of these blocks.
}
\label{member}
\end{figure}

\begin{figure}
\begin{center}
\includegraphics[width = \textwidth]{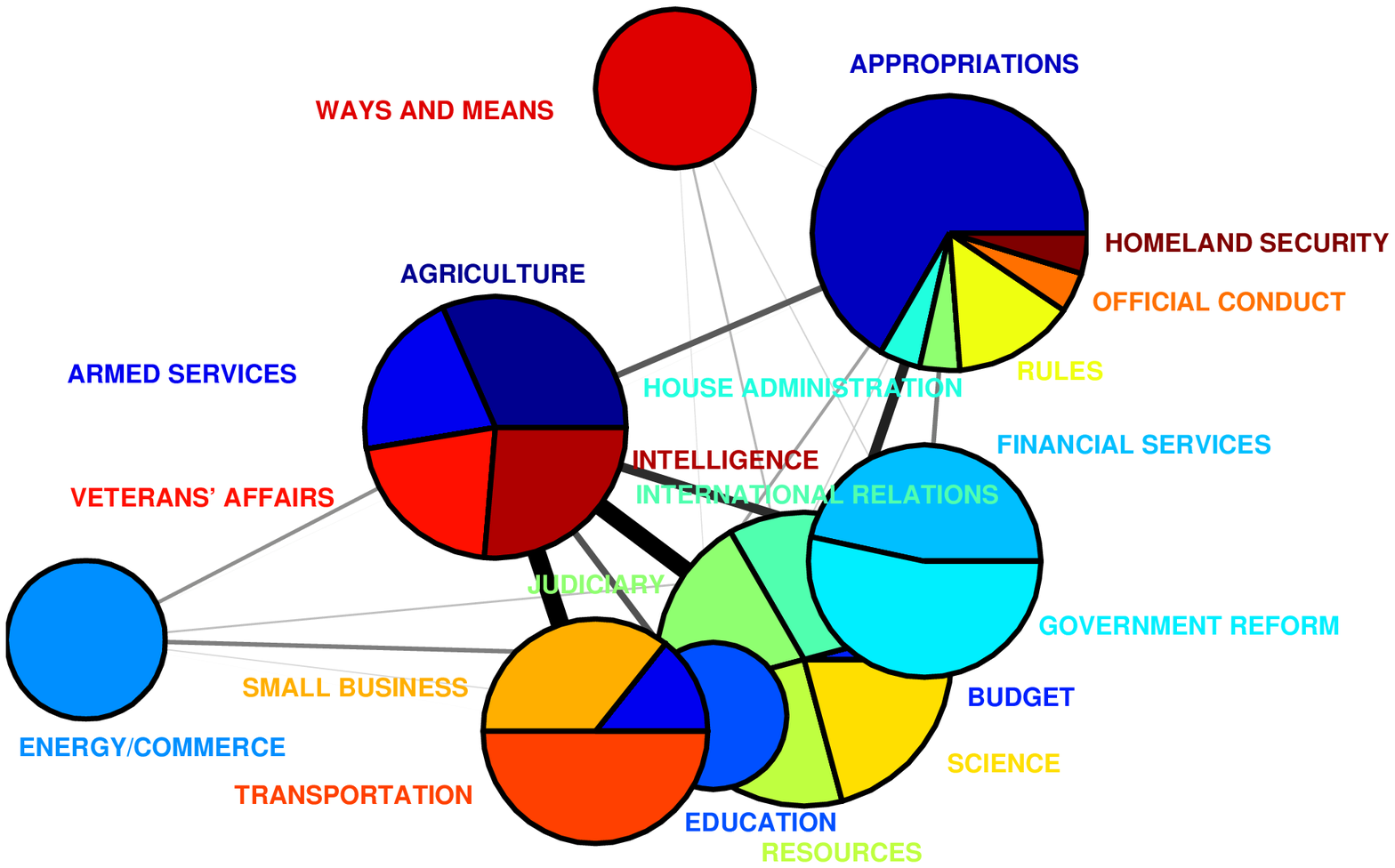}
\end{center}
\caption{(Color) Pie chart of the 107th House
  at the modularity-maximizing
  organizational level indicated by the dashed dividing ring in
  Fig.~\ref{modcolor}.  Each pie represents a community and has wedges colored by the
  parent standing and select committees of the (sub)committees present in that community.  The size of a wedge is determined by the number of (sub)committees it contains.  Only connections between different communities are depicted (with thicker lines indicating stronger connections); the intra-community edges are not visible at this level of organization.  
  }
\label{piechart}
\end{figure}

Using a ``membership matrix'' that encodes this combined information (see
Fig.~\ref{member}), we manipulate the resulting collection of local
communities to produce dendrograms (this procedure is described in
Ref.~\cite{bagrow}).  An unsorted membership matrix $N$ collects the
ensemble of information about the local communities of each node as
originally ordered in the data. Because each subcommittee is listed with
its parent committee in the data ordering, this unsorted membership
matrix (Fig.~\ref{member}a) is already nearly block partitioned.  The $j$th
element of the $i$th row has the value 1 if node $j$ is part of $i$'s
community and the value 0 if not.  We compare the values in two rows ($i$
and $k$) and define a distance $\Delta$ between them according to the
number of times they differ:
\begin{equation}
	\Delta(i,l) = n - \sum_{l = 1}^n \delta(N_{il},N_{kl})\,,
\end{equation}
where 
$\delta$ is 1 if $N_{il} = N_{kl}$ and 0 if not.  As discussed in
Ref.~\cite{bagrow}, we then compute a sorted membership matrix $\tilde{N}$
as follows: (1) compute the distance $\Delta(i,l)$ for all rows $l > i$;
(2) swap the row $i + 1$ with the row $i_{\Delta}$ that has the smallest
distance to row $i$ (equivalent to interchanging vertex labels, so columns
$i$ and $i_\Delta$ must also be swapped); (3) repeat steps 1 and 2 for row
$i + 1$ and continue until there are no remaining rows to consider.
Well-separated communities in $\tilde{N}$ appear as blocks along the
diagonal, and their imperfections indicate possible substructure [compare
panels (a) and (b) of Fig.~\ref{member}].

We obtain well-structured dendrograms over a wide range of values of the
local clustering threshold parameter $\alpha$ and again observe the close
connections between the Rules and Homeland Security committees in the 107th
House (see, for example, the 7 o'clock position in Fig.~\ref{modcolor}).
These ties between the Rules and Homeland Security committees are evident
even for values of $\alpha$ for which some of the other groupings in the
dendrogram have disappeared, again indicating the strength of their
connection.

One can depict the network's communities (and how strongly they are
connected to each other) at a given level of organization using pie charts
(which provide a coarse-graining of the network reminiscent of the
``cartographic" visualization of networks discussed in \cite{amaral}). For
example, Fig.~\ref{piechart} shows such a pie chart for the 107th House, with each pie representing a community and the color of each wedge indicating the parent
standing or select committee of the (sub)committees therein. More generally, it is desirable to
study not only separate communities in networks but also the overlap
between such communities and what role that can play in the transfer of
information and ideas.  Such considerations have the potential to be very
interesting, as committees with strong ties to multiple communities may
have a substantial level of influence with otherwise disparate groups.
While practically every study of community structure in networks ignores
community overlap \textit{a priori}, a few researchers have begun to
scrutinize this feature (see, for example, \cite{vicsek}).  The methods
that have been employed in such studies are different from those discussed
here, but one can compute basic statistics from the membership matrices to
get some indication of community overlap.  For example, the mean community
size in the 107th House at the organizational level that gives the highest
maximum modularity is 11.78 nodes and the standard deviation is 3.37 nodes.
(The results from other Congresses are summarized in Table~\ref{overlap}.)
By comparison, the set of communities at this organizational level (see
Fig.~\ref{piechart}) has only 8 communities, giving an average community
size of $113/8 \approx 14.1$ committees per community.  The average
standing committee contains $113/21 \approx 5.4$ subcommittees (counting
the parent committee).  Community sizes vary roughly with the threshold
parameter $\alpha$ (which is selected to give the highest maximum
modularity), with smaller values of $\alpha$ typically yielding larger communities by
construction.  Using the 107th House as an example, we can see from
Table~\ref{overlap} and the combination of Figs.~\ref{modcolor},
\ref{member}, and \ref{piechart} that this weighted local community
detection seemingly indicates a relatively small amount of overlap between the
locally-defined communities.

\begin{table}
\centerline{
\begin{tabular}{|c|c|c|c|} \hline
House & Mean & Standard Deviation & Local Clustering Threshold \\ \hline
101st & 8.9325 & 3.9785 & 0.1595  \\ \hline
102nd & 20.3558 & 5.1660 & 0.08095 \\ \hline
103rd & 17.0567 & 6.0884 & 0.09895 \\ \hline
104th & 10.8962 & 4.2537 & 0.1131 \\ \hline
105th & 7.4259 & 2.9425 & 0.1640 \\ \hline
106th & 46.5888 & 6.6118 &  0.04582 \\ \hline
107th & 11.7788 & 3.3746 &  0.1179 \\ \hline
108th & 8.8136 & 2.6212 & 0.1490 \\ \hline
\end{tabular}}\caption{Community sizes in the 101st--108th Houses at the organizational level determined by the local clustering threshold value $\alpha$ that gives the highest maximum modularity.  For each Congress, we list the mean number of nodes, the standard deviation, and the value of $\alpha$.}\label{overlap}
\end{table}

\subsection{Direct comparison of dendrograms}

In Table~\ref{modcompare}, we list for each of our methods the maximum modularity obtained for a single ``cut" through the dendrogram and the average modularity over all possible cuts.  A cut signifies an organizational level of a dendrogram; we depict a cut graphically using a concentric circular ring of the appropriate radius that divides inter-community links (those outside the ring) from intra-community ones.  See, for example, the dendrogram in Fig.~\ref{modcolor} and the resulting community-composition pie chart in Fig.~\ref{piechart}.  For the weighted local community
detection method, we used the values of the local clustering threshold
$\alpha$ (denoted $\alpha_1\,, \dots \,, \alpha_8$ for the 101st--108th
Houses) giving the dendrograms with highest maximum modularity.  These
values are given in Table~\ref{overlap}.
Similar modularity values are obtained over a relatively broad range
of~$\alpha$.  To see the number of organizational levels revealed by each
algorithm, we list in Table~\ref{HStable} the maximum, mean, and standard
deviation of the local Strahler numbers for the dendrograms produced for
each House.  Observe, for example, that the weighted local community
detection method finds the largest number of organizational levels.

\begin{table}
\centerline{
\begin{tabular}{|c|c|c|c|c|} \hline
Method: & SL & EB & NB & WL \\ \hline
$S_{1}\,,\,\tilde{S}_{1}\,,\,S_{\sigma_1}$ & 4\,, 3.289\,, 0.758 &
3\,, 2.516\,, 0.501 & 4\,, 2.786\,, 0.799 & 5\,, 4.638\,, 0.552 \\
\hline $S_{2}\,,\,\tilde{S}_{2}\,,\,S_{\sigma_2}$ & 4\,, 3.203\,,
0.686 & 4\,, 2.620\,, 0.669 & 3\,, 2.350\,, 0.478 & 5\,, 4.522\,,
0.695 \\ \hline $S_{3}\,,\,\tilde{S}_{3}\,,\,S_{\sigma_3}$ & 4\,,
3.362\,, 0.679 & 4\,, 3.064\,, 0.872 & 3\,, 2.582\,, 0.495 & 5\,,
4.567\,, 0.645 \\ \hline $S_{4}\,,\,\tilde{S}_{4}\,,\,S_{\sigma_4}$
& 5\,, 4.547\,, 0.770 & 3\,, 2.613\,, 0.489 & 3\,, 2.604\,, 0.491 &
5\,, 4.368\,, 0.588 \\ \hline
$S_{5}\,,\,\tilde{S}_{5}\,,\,S_{\sigma_5}$ & 5\,, 3.880\,, 0.924 &
3\,, 2.546\,, 0.500 & 3\,, 2.500\,, 0.502 & 5\,, 4.639\,, 0.585 \\
\hline $S_{6}\,,\,\tilde{S}_{6}\,,\,S_{\sigma_6}$ & 4\,, 3.626\,,
0.575 & 4\,, 3.393\,, 0.491 & 3\,, 2.486\,, 0.502 & 5\,, 4.243\,,
0.609 \\ \hline $S_{7}\,,\,\tilde{S}_{7}\,,\,S_{\sigma_7}$ & 5\,,
4.089\,, 0.819 & 4\,, 2.779\,, 0.799 & 3\,, 2.575\,, 0.497 & 5\,,
4.699\,, 0.459 \\ \hline $S_{8}\,,\,\tilde{S}_{8}\,,\,S_{\sigma_8}$
& 5\,, 4.509\,, 0.767 & 3\,, 2.466\,, 0.499 & 3\,, 2.492\,, 0.500 &
5\,, 4.254\,, 0.666 \\ \hline
\end{tabular}}
\caption{Horton-Strahler numbers ($S_i$), mean local Horton-Strahler
numbers ($\tilde{S}_i$), and the standard deviation ($\sigma_i$) of the
local Horton-Strahler numbers for the 101st--108th Houses ($i = 1$ denotes
the 101st House of Representatives, etc.), with dendrograms produced using single linkage clustering (SL), random-walk betweenness with sequential edge (committee assignment) removal (EB), random-walk betweenness with sequential Representative node removal (NB), and a
weighted generalization (WL) of the local community detection method
of~\cite{bagrow} with the $\alpha$ values that give the highest maximum
modularity.}\label{HStable}
\end{table}

We compare dendrograms at the cuts (organizational levels)
corresponding to their respective maximum modularities.
Table~\ref{compare} collects these comparisons across the different
clustering algorithms considered for each of the 101st--108th Houses.  We
compare the algorithms in pairs, with each tabulated entry indicating the
fraction of committee pairs classified in the same manner by both methods
(that is, both methods identify the committee pair as belonging to the same
community or both methods identify the pair as belonging to separate
communities).  Although we list the results from specific
maximum-modularity cuts in Table~\ref{compare}, we obtained similar values
over broad ranges of cuts in the dendrograms.

\begin{table}
\centerline{
\begin{tabular}{|c|c|c|c|c|c|c|} \hline
House & WL vs SL & WL vs EB & WL vs NB & SL vs EB & SL vs NB & EB vs NB \\ \hline
101st & 0.8056 & 0.9123 & 0.7813 & 0.7630 & 0.6077 & 0.7613 \\ \hline
102nd & 0.8349 & 0.4876 & 0.5837 & 0.3988 & 0.5000 & 0.8306 \\ \hline
103rd & 0.7896 & 0.5463 & 0.7046 & 0.4751 & 0.6301 & 0.7289 \\ \hline
104th & 0.8794 & 0.6961 & 0.7191 & 0.6453 & 0.6855 & 0.8767 \\ \hline
105th & 0.8927 & 0.6433 & 0.4962 & 0.6274 & 0.4886 & 0.8221 \\ \hline
106th & 0.7907 & 0.6762 & 0.6662 & 0.6708 & 0.5860 & 0.7988 \\ \hline
107th & 0.8685 & 0.6775 & 0.5397 & 0.6841 & 0.5653 & 0.8274 \\ \hline
108th & 0.8893 & 0.6975 & 0.5950 & 0.6407 & 0.5196 & 0.7876 \\ \hline
\end{tabular}}\caption{Comparison of House community structure as
identified using different algorithms for the 101st--108th Congresses.  The
numbers indicate the fraction of leaf pairs identified in the same manner
in pairwise comparisons of single linkage clustering (SL), edge betweenness
(EB), node betweenness (NB), and the weighted local community detection
method with maximum modularity (WL).  Leaf pairs are identified in the same
manner in two dendrograms if, at a given organizational level, both dendrograms place them in the same subtree or both place them in different
subtrees.  For each House, we use the organizational level identified as
having the highest maximum modularity.  We obtain similar comparison values
over broad ranges of cuts in the dendrograms.}\label{compare}
\end{table}

To illustrate these results, we compare the similarity scores in
Table~\ref{compare} to the dendrograms for the 107th House produced using
single linkage clustering (Fig.~\ref{107clust}), betweenness-based edge
removal (Fig.~\ref{107biedge}), and our local community detection method
(Fig.~\ref{modcolor}), as well as the single linkage clustering dendrogram
for the 108th House (Fig.~\ref{108hs}).  The maximum-modularity cuts in
these dendrograms have 28, 5, 8, and 25 communities, respectively.  Several
observations are evident from Table~\ref{compare}.  For example, the
betweenness-based algorithms produce results that are quantitatively
similar to each other but in general less similar to the other two methods.
The weighted local clustering method and single linkage clustering likewise
produce similar community structures.

Even when the quantitative measure of community similarity at the preferred
cuts is low, many committees of interest nevertheless get grouped similarly
in dendrograms produced from multiple methods, suggesting that the observed
close ties between these committees are properties of the networks
themselves rather than of the algorithms used.  For example, the Select
Committee on Homeland Security of the 107th House is grouped with the Rules
Committee and its subcommittees using single linkage clustering
(Fig.~\ref{107clust}), the weighted local community determination method
(Fig.~\ref{modcolor}), and the edge-betweenness based method
(Fig.~\ref{107biedge}).  One can also see that the Permanent Select
Committee on Intelligence and its subcommittees are grouped together by all
three algorithms.  The recurrence of such groupings in the dendrograms, and
in the visualizations of Figs.~\ref{network}--\ref{network108}, further
supports the claim that these connections are inherent properties of the
networks themselves.


\end{document}